\documentclass[12pt]{article}
\pdfoutput=1

\usepackage{amsmath,amssymb}
\usepackage{bm}
\usepackage{color}
\usepackage{graphicx}

\newcommand{\Slash}[1]{{\ooalign{\hfil/\hfil\crcr\(#1\)}}}

\begin{document}
\title{
\begin{flushright}
\ \\*[-80pt]
\begin{minipage}{0.2\linewidth}
\normalsize
HUPD1714 \\*[10pt]
\end{minipage}
\end{flushright}
{\Large \bf
Effective theory analysis for vector-like quark model
\\*[10pt]}}

\author{
\centerline{
Takuya~Morozumi$^{1,2}$\footnote{E-mail address: morozumi@hiroshima-u.ac.jp},
~Yusuke~Shimizu$^{1,}$\footnote{E-mail address: yu-shimizu@hiroshima-u.ac.jp},
}\\
\centerline{
~Shunya~Takahashi$^{1,}$\footnote{E-mail address: s-takahashi@hiroshima-u.ac.jp},~and~
Hiroyuki~Umeeda$^{3,}$\footnote{E-mail address: umeeda@riko.shimane-u.ac.jp}}
\\*[12pt]
\centerline{
\begin{minipage}{\linewidth}
\begin{center}
$^1${\it \normalsize
Graduate~School~of~Science,~Hiroshima~University, \\
Higashi-Hiroshima~739-8526,~Japan} \\*[5pt]
$^2${\it \normalsize
Core~of~Research~for~the~Energetic~Universe,~Hiroshima~University, \\
Higashi-Hiroshima~739-8526,~Japan} \\*[5pt]
$^3${\it \normalsize
Graduate School of Science and Engineering, Shimane University, \\
Matsue 690-8504, Japan}
\end{center}
\end{minipage}}
\\*[50pt]}

\date{
\centerline{\small \bf Abstract}
\begin{minipage}{0.9\linewidth}
\medskip
\medskip
\small
We study a model with a down-type SU(2) singlet vector-like quark (VLQ) as a minimal extension of the standard model (SM).
In this model, flavor changing neutral currents (FCNCs) arise at tree level
and the unitarity of the $3\times 3$ Cabibbo-Kobayashi-Maskawa (CKM) matrix does not hold.
In this paper, we constrain the FCNC coupling from $b\rightarrow s$ transitions, especially $B_s\rightarrow \mu^+\mu^-$ and $\bar{B}\rightarrow X_s\gamma$ processes.
In order to analyze these processes,
we derive an effective Lagrangian which is valid below the electroweak symmetry breaking scale.
For this purpose, we first integrate out the VLQ field and derive an effective theory by matching Wilson coefficients up to one-loop level.
Using the effective theory, we construct the effective Lagrangian for $b\rightarrow s\gamma^{(*)}$.
It includes the effects of the SM quarks and the violation of the CKM unitarity.
We show the constraints on the magnitude of the FCNC coupling and its phase by taking account of the current experimental data on $\Delta M_{B_s}$, $\mathrm{Br}[B_s\rightarrow\mu^+\mu^-]$, $\mathrm{Br}[\bar{B}\rightarrow X_s\gamma]$ and CKM matrix elements as well as theoretical uncertainties.
We find that the constraint from the $\mathrm{Br}[B_s\rightarrow\mu^+\mu^-]$ is more stringent than that from the $\mathrm{Br}[\bar{B}\rightarrow X_s\gamma$].
We also obtain the bound for the mass of the VLQ and the strength of the Yukawa couplings related to the FCNC coupling of $b\rightarrow s$ transition.
Using the CKM elements which satisfy above constraints,
we show how the unitarity is violated on the complex plane.
\end{minipage}
}

\begin{titlepage}
\maketitle
\thispagestyle{empty}
\end{titlepage}

\section{Introduction}
After the discovery of the Glashow-Iliopoulos-Maiani (GIM) mechanism \cite{Glashow:1970gm},
this suppression mechanism of the flavor changing neutral
current (FCNC) is firmly verified in $K$, $D$ and $B$ meson systems.
The unitarity of the Cabibbo-Kobayashi-Maskawa (CKM) matrix \cite{Kobayashi:1973fv} is also verified.
As investigated in Refs.\ \cite{Charles:2004jd, Bona:2007vi},
the CKM unitarity is consistent with current data,
which characterizes one of the most successful aspects in the standard model (SM).
\par
As an extension of the quark sector,
vector-like quark (VLQ) is considered.
Here, VLQ is a quark whose
representations in gauge group
for left- and right-handed components
are the same.
As models including VLQs, some new physics scenarios
are considered in the literature.
Such vector-like extensions of the SM
include the universal seesaw model \cite{Berezhiani:1983hm}.
This scenario introduces gauge singlet vector-like fermions to
explain hierarchical structure of fermion masses.
Furthermore,
in the context of left-right symmetry,
the seesaw mechanism induced by vector-like fermions
gives solution to the strong CP problem \cite{Babu:1989rb}.
\par
The model with VLQ leads to the rich phenomenology which can be testable in experiments~\cite{Barenboim:2001fd}-\cite{Bobeth:2016llm}.
In particular, FCNCs induced by VLQ
give rise to deviation from the SM prediction.
Furthermore, the unitary relation of the CKM matrix, {\it e.g.},
$V_{ub}^*V_{us}+V_{cb}^*V_{cs}+V_{tb}^*V_{ts}=0$,
no longer holds.
The unitarity triangle is modified as a quadrangle
due to correction which arises from FCNCs.
On the other hand, the direct detection of the VLQ
is under way in collider experiments \cite{Aad:2015kqa}.
Then, the prediction and constraint on the mass and couplings
of VLQ from the flavor observables provide them with important information.
\par
In this paper, a model including one additional down-type VLQ is discussed.
Integrating out VLQ, one can find that tree level
FCNC arises from interaction with $Z$ boson
and Higgs bosons.
On the basis of the effective field theory (EFT), we derive loop functions
which correspond to the Inami-Lim functions in the SM.
In order to examine the FCNC,
phenomenological analysis is carried out for $b\to s$ transition.
Specifically, experimental data of $\bar{B}\to X_s\gamma$, $B_s\to\mu^+\mu^-$ and
the mass difference in $B_s-\bar{B_s}$ system
are utilized to constrain the model.
The constraints on the magnitude of the FCNC coupling and its phase are shown
by taking account of the current experimental data as well as theoretical uncertainties.
\par
This paper is organized as follows:
In Sec.~\ref{sec:2}, we integrate out down-type VLQ
and determine Wilson coefficients of the EFT up to one-loop level.
Loop functions
are summarized in Sec.~\ref{sec:3}.
In Sec.~\ref{sec:4}, the phenomenological analysis for $b\to s$ transition is given.
Section~\ref{sec:5} is devoted to summary and discussion.

\section{Integrating out VLQ fields}
\label{sec:2}
In this section, we derive a low energy effective Lagrangian by integrating out VLQ fields. For this purpose, we show a full Lagrangian which includes one down-type SU(2) singlet VLQ in addition to the SM quarks. We assume that the mass of the VLQ is much larger than the electroweak (EW) scale. Then the Lagrangian $\mathcal{L_\mathrm{Full}}$ which is invariant under $\mathrm{SU(3)_c}\times \mathrm{SU(2)}\times \mathrm{U(1)_Y}$ is
\begin{align}
  \mathcal{L}_{\mathrm{Full}}&=
  \overline{q^i_L}i\Slash{D}_Lq_L^i+\overline{u_R^i}i\Slash{D}_R^uu_R^i+\overline{d_R^i}i\Slash{D}_R^dd_R^i+\overline{d_L^4}i\Slash{D}_R^dd_L^4+\overline{d_R^4}i\Slash{D}_R^dd_R^4\nonumber\\[4pt]
  &\phantom{=}-
  [y_d^{ij}\overline{q_L^i}\phi d_R^j
  +y_d^{i4}\overline{q_L^i}\phi d_R^4
  +M_4\overline{d_L^4}d_R^4+y_u^{ii}\overline{u_R^i}\tilde{\phi}q_L^i+\mathrm{h.c.}]~,
\label{eq:FullTheory}
\end{align}
where $i=1,2,3$ denotes the indices for generations. $d_{L,R}^4$ are VLQs. $y_{u}$ and $y_d$ represent Yukawa couplings of up-type and down-type quarks respectively. The matrix for Yukawa coupling of up-type quarks is taken to be real diagonal, while that of down-type quarks is a $3\times 4$ matrix. $M_4$ denotes the mass of VLQ.
Note that the mixing term between the left-handed VLQ $d^4_L$ and right-handed SM down-type quarks $d^i_R$ is allowed in general. However we can remove the mixing term by the rotation of the down-type quarks. Hence we can take the Lagrangian as Eq.~\eqref{eq:FullTheory}.
The covariant derivatives are defined as follows:
\begin{align}
D_{L\mu}&=\partial_\mu + ig_s\frac{\lambda^a}{2}G_\mu^a + ig\frac{\tau^I}{2}W_\mu^I+ig'\frac{Y_{qL}}{2}B_\mu ~,\label{Eq:cov_qL}\\
D_{R\mu}^u&=\partial_\mu + ig_s\frac{\lambda^a}{2}G_\mu^a + ig'\frac{Y_{uR}}{2}B_\mu ~,\label{Eq:cov_uR}\\
D_{R\mu}^d&=\partial_\mu + ig_s\frac{\lambda^a}{2}G_\mu^a + ig'\frac{Y_{dR}}{2}B_\mu ~,
\label{Eq:cov_dR}
\end{align}
where $\lambda^a,~\tau^I$ and $Y_X$ are Gell-Mann matrices, Pauli matrices and $\mathrm{U(1)_Y}$ hypercharge of a field $X$ $(X=q_L,u_R,d_R)$ respectively.

\subsection{Matching full theory and effective theory}
\begin{figure}[tbp]
  \centering
  \includegraphics[width=14.0cm]{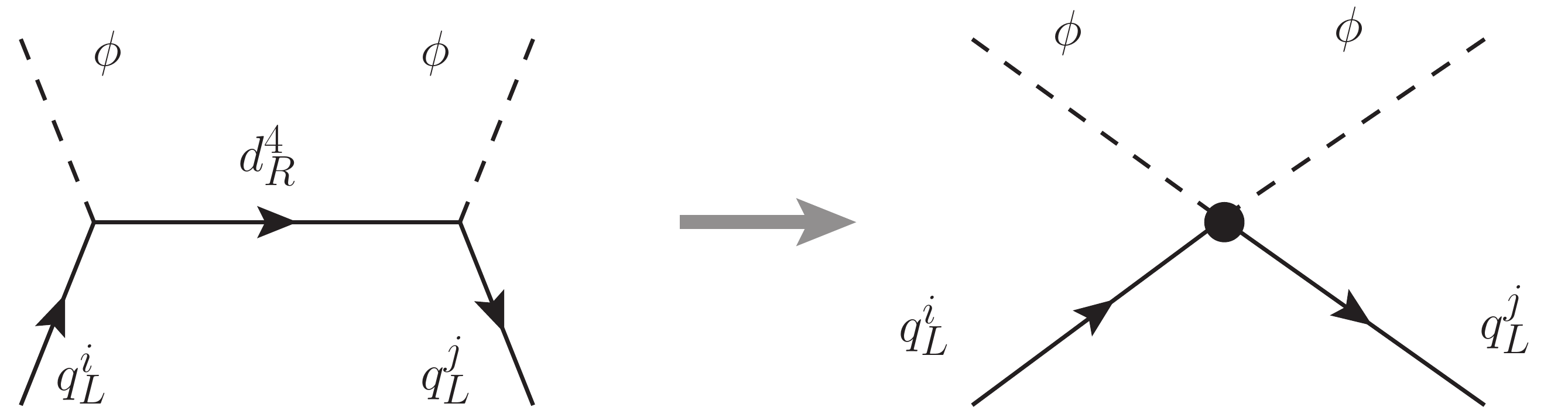}
  \caption{The Feynman diagrams for scattering of a pair of quark and anti-quark into a Higgs pair $(q^i\overline{q}^j\rightarrow\phi\phi^\dagger)$. The left figure shows the diagram of the full theory in which the VLQ is exchanged, while the right figure shows the diagram of the effective theory where the VLQ is absent and already integrated out.}
  \label{Fig:TreeLevelMatching}
\end{figure}

In order to obtain the higher dimensional operators which represent the effect of the VLQ in the energy scale between $M_4$ and the EW scale, we integrate out the VLQ fields $d^4_{L,R}$ in Eq.~\eqref{eq:FullTheory}. At first, we perform tree level matching at VLQ mass scale $M_4$. In  Fig.~~\ref{Fig:TreeLevelMatching}, we show the Feynman diagram (left figure) for scattering of a pair of quark and anti-quark into a Higgs pair $(q^i\overline{q}^j\rightarrow\phi\phi^\dagger)$, in which the VLQ is exchanged.
We assume that the external particles have momenta much smaller than the mass of the VLQ. Then the amplitude of the left figure can be reproduced up to $\mathcal{O}(M_4^{-2})$ accuracy by computing the Feynman diagram of the right figure with the following low energy effective Lagrangian \cite{Ishiwata:2015cga,Bobeth:2016llm,delAguila:2000aa,delAguila:2000rc,Chen:2017hak},
\begin{align}
  \mathcal{L}_{\mathrm{Eff}}^{\mathrm{tree}}=
  i\frac{y_d^{j4}y_d^{i4*}}{M_4^2}\left(\overline{q_L^j}\phi\right)\Slash{D}_R^d
  \left(
  \phi^\dag q_L^i
  \right)~,
  \label{Eq:TreeOperator}
\end{align}
where $i,j=1,2,3$.
The effective Lagrangian is written in terms of dimension-six operator and its coefficient is determined so that it reproduces the amplitude of the left figure in Fig.~\ref{Fig:TreeLevelMatching} within the precision of $\mathcal{O}(M_4^{-2})$. By using the equation of motions derived from SM Lagrangian, we can rewrite the effective Lagrangian of Eq.~\eqref{Eq:TreeOperator},
\begin{align}
  \mathcal{L}_{\mathrm{Eff}}^\mathrm{tree}
  &=
  \frac{y_d^{j4}y_d^{i4*}}{2M_4^2}
  \left[
  \frac{i}{2}\left(\overline{q_L^j}\tau^I\gamma^\mu q_L^i\right)
  \left\{
  \left(D_\mu\phi\right)^\dag\tau^I\phi
  -
  \phi^\dag\tau^I\left(D_\mu\phi\right)
  \right\}\right.\nonumber\\[3pt]
  &\phantom{=}\left.
  +\frac{i}{2}\left(\overline{q_L^j}\gamma^\mu q_L^i\right)
  \left\{
  \left(D_\mu\phi\right)^\dag\phi
  -
  \phi^\dag\left(D_\mu\phi\right)
  \right\}\right.\nonumber\\[3pt]
  &\phantom{=}\left.
  +\left(\phi^\dag\phi\right)
  \left(
  y_d^{ik}\overline{q_L^j}\phi d_R^k+y_d^{jk*}\overline{d_R^k}\phi^\dag q_L^i
  \right)\right.\nonumber\\[3pt]
  &\phantom{=}\left.
  +\frac{1}{2}\left(\phi^\dag\tau^I\phi\right)
  \left\{
  y_u^{jj}\left(\overline{q_L^j}\tau^I\tilde{\phi}\right)u_R^j
  +y_u^{jj*}\overline{u_R^j}\left(\tilde{\phi}^\dag\tau^I q_L^j\right)
  \right\}\right.\nonumber\\[3pt]
  &\phantom{=}\left.
  +\frac{1}{2}\left(\phi^\dag\phi\right)
  \left\{
  y_u^{jj}\overline{q_L^j}\tilde{\phi}u_R^j+y_u^{jj*}\overline{u_R^j}\tilde{\phi}^\dag q_L^j
  \right\}
  \right]~.
  \label{Eq:TreeOperator2}
\end{align}

\begin{figure}[tbp]
  \centering
  \includegraphics[width=14.0cm]{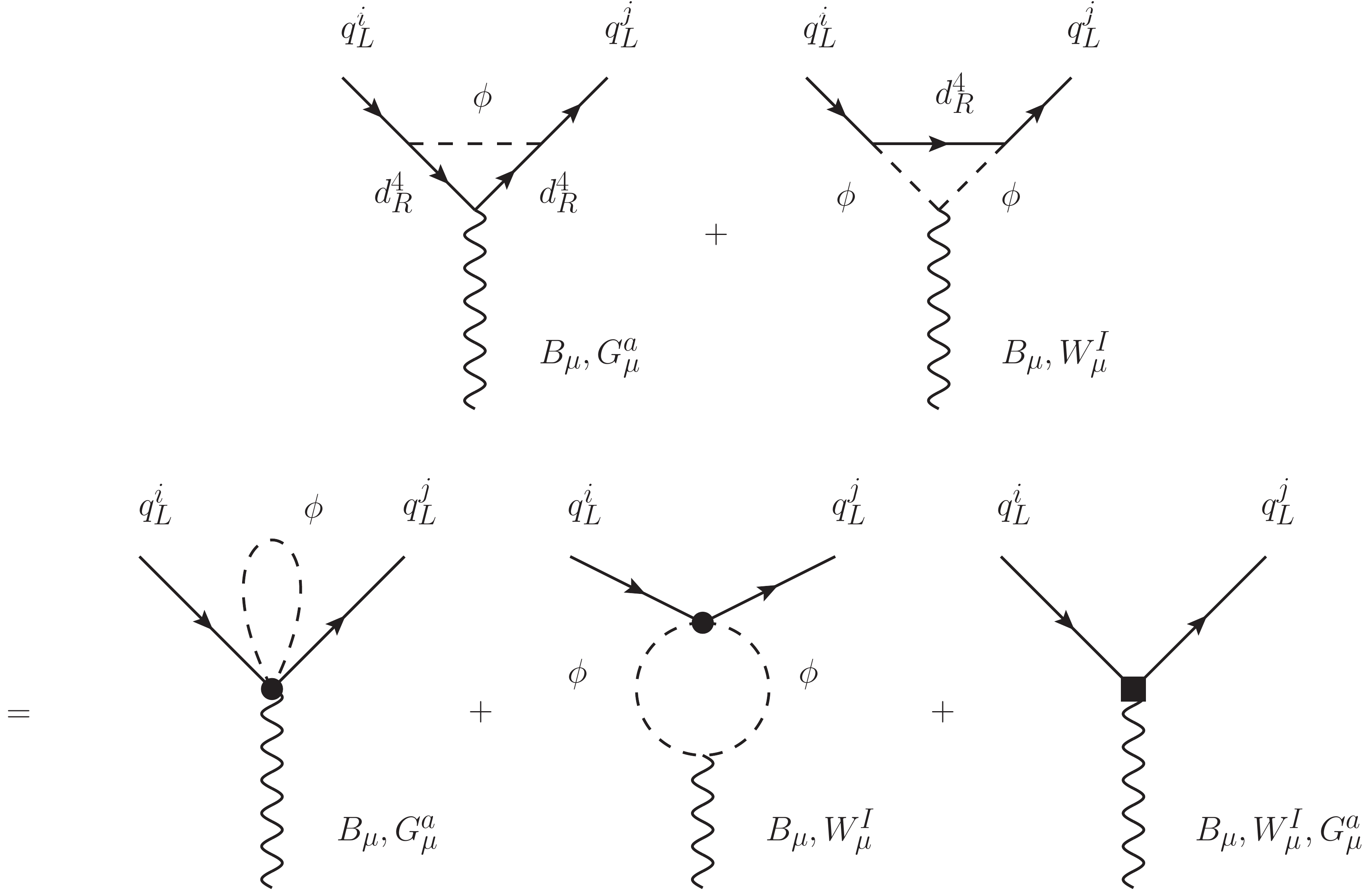}
  \caption{The Feynman diagrams for the decay of $q^i_L$ into $q^j_L$ and one of gauge fields $B$, $W^I$ or $G^a$ at one-loop level with the full theory (top figures) and the effective theory (bottom-left and bottom-center figures). The circular marks denote the tree level effective operator and the square mark denotes the new effective operators.}
  \label{Fig:LoopMatching}
\end{figure}

Next we consider one-loop level matching between the full theory and the effective theory to obtain effective interactions which contribute to the radiative transition of the quarks. The procedure is as follows:
\begin{description}
  \item[\hspace{0.3cm}(i)] We calculate the amplitudes of the Feynman diagrams for the decay of $q^i_L$ into $q^j_L$ and one of gauge fields $B$, $W^I$ or $G^a$ at one-loop level (See the top figures in Fig.~\ref{Fig:LoopMatching}.). These diagrams include the VLQ in the internal line. In this calculation, we renormalize the amplitudes with the $\mathrm{\overline{MS}}$ scheme.
  \item[\hspace{0.3cm}(ii)] We calculate the same transitions as those of the procedure (i) with the effective operator in Eq.~\eqref{Eq:TreeOperator} obtained by tree level matching (See the bottom-left and bottom-center figures in Fig.~\ref{Fig:LoopMatching}.). In this calculation, we also renormalize the amplitudes with the $\mathrm{\overline{MS}}$ scheme.
  \item[\hspace{0.3cm}(iii)] We introduce new effective operators and determine their coefficients so that the renormalized amplitudes in procedure (ii) match with those of the full theory computed in procedure (i) (See the bottom-right figure in Fig.~\ref{Fig:LoopMatching}.).
\end{description}
We can obtain the following effective Lagrangian $\mathcal{L}_\mathrm{Eff}^\mathrm{one-loop}$ at one-loop level:
\begin{align}
  \mathcal{L}_\mathrm{Eff}^\mathrm{one-loop}
  =
  \mathcal{L}_\mathrm{Eff}^K
  +
  \mathcal{L}_\mathrm{Eff}^B
  +
  \mathcal{L}_\mathrm{Eff}^W
  +
  \mathcal{L}_\mathrm{Eff}^G~,
\end{align}
where
\begin{align}
  \mathcal{L}_\mathrm{Eff}^K
  &=
  i\frac{y_d^{j4}y_d^{i4*}}{16\pi^2}
  \left(
  \frac{1}{2}\ln\frac{\mu^2}{M_4^2}+\frac{3}{4}
  \right)\overline{q_L^j}\Slash{D}_Lq_L^i\nonumber\\[3pt]
  &+
  i\frac{1}{16\pi^2}\frac{y_d^{j4}y_d^{i4*}}{3M_4^2}
  \left(
  y_d^{jl*}\overline{d_R^l}\phi^\dagger + y_u^{jj*}\overline{u_R^j}\tilde{\phi}^\dagger
  \right)\Slash{D}_L
  \left(
  y_d^{ik}\phi d_R^k + y_u^{ii}\tilde{\phi}u_R^i
  \right)~,\label{Eq:OneLoopOpe}
\end{align}
\begin{align}
  \mathcal{L}_\mathrm{Eff}^B
  &=
  g'^2\frac{y_d^{j4}y_d^{i4*}}{16\pi^2M_4^2}
  \left\{
  \frac{Y_{dR}}{2}\cdot\frac{7}{36}
  -
  \frac{Y_\phi}{2}\left(
  \frac{1}{6}\ln\frac{\mu^2}{M_4^2}+\frac{11}{36}
  \right)
  \right\}\nonumber\\[3pt]
  &\phantom{=}\times\left(\overline{q_L^j}\gamma^\mu q_L^i\right)
  \left[
  \frac{Y_{lL}}{2}\overline{l_L}\gamma_\mu l_L + \frac{Y_{eR}}{2}\overline{e_R}\gamma_\mu e_R
  +\frac{Y_{qL}}{2}\overline{q_L}\gamma_\mu q_L + \frac{Y_{uR}}{2}\overline{u_R}\gamma_\mu u_R\right.\nonumber\\[3pt]
  &\phantom{=}\left.
  +\frac{Y_{dR}}{2}\overline{d_R}\gamma_\mu d_R + i\frac{Y_\phi}{2}
  \left\{
  \phi^\dagger D_\mu\phi-(D_\mu\phi)^\dagger\phi
  \right\}
  \right]\nonumber\\[3pt]
  &\phantom{=}
  +g'\frac{1}{16\pi^2M_4^2}
  \left(
  \frac{Y_{qL}}{2}\cdot\frac{1}{12}-\frac{Y_{dR}}{2}\cdot\frac{1}{8}
  \right)\left[
  y_d^{j4}y_d^{i4*}\overline{q_L^j}\sigma_{\mu\nu}
  \left(
  y_d^{il}\phi d_R^l + y_u^{ii}\tilde{\phi}u_R^i
  \right)B^{\mu\nu} + \mathrm{h.c.}
  \right]~,\label{Eq:OneLoopOpe_B}
\end{align}
\begin{align}
  \mathcal{L}_{\mathrm{Eff}}^W
  &=
  -g^2\frac{y_d^{j4}y_d^{i4*}}{16\pi^2M_4^2}
  \left(
  \frac{1}{6}\ln\frac{\mu^2}{M_4^2} + \frac{11}{36}
  \right)\overline{q_L^j}\gamma^\mu\frac{\tau^I}{2}q_L^i\nonumber\\[3pt]
  &\phantom{=}\times
  \left[
  \overline{l_L}\frac{\tau^I}{2}\gamma_\mu l_L + \overline{q_L}\frac{\tau^I}{2}\gamma_\mu q_L + i\left\{
  \phi^\dagger\frac{\tau^I}{2}D_\mu\phi - (D_\mu\phi)^\dagger\frac{\tau^I}{2}\phi
  \right\}
  \right]\nonumber\\[3pt]
  &\phantom{=}
  +g\frac{1}{16\pi^2M_4^2}\cdot\frac{1}{12}
  \left\{
  y_d^{j4}y_d^{i4*}\overline{q_L^j}\frac{\tau^I}{2}\sigma_{\mu\nu}
  \left(
  y_d^{il}\phi d_R^l + y_u^{ii}\tilde{\phi}u_R^i
  \right)W^{I\mu\nu} + \mathrm{h.c.}
  \right\}~,\label{Eq:OneLoopOpe_W}
\end{align}
\begin{align}
  \mathcal{L}_\mathrm{Eff}^G
  &=
  g_s^2\frac{y_d^{j4}y_d^{i4*}}{16\pi^2M_4^2}\cdot\frac{7}{36}
  \left(
  \overline{q_L^j}\gamma^\mu\frac{\lambda^a}{2}q_L^i
  \right)
  \left(
  \overline{q_L}\frac{\lambda^a}{2}\gamma_\mu q_L
  +\overline{u_R}\frac{\lambda^a}{2}\gamma_\mu u_R
  +\overline{d_R}\frac{\lambda^a}{2}\gamma_\mu d_R
  \right)\nonumber\\[3pt]
  &\phantom{=}
  +g_s\frac{1}{16\pi^2M_4^2}\left(
  -\frac{1}{24}
  \right)
  \left\{
  y_d^{j4}y_d^{i4*}\overline{q_L^j}\frac{\lambda^a}{2}\sigma_{\mu\nu}
  \left(
  y_d^{il}\phi d_R^l + y_u^{ii}\tilde{\phi}u_R^i
  \right)G^{a\mu\nu} + \mathrm{h.c.}
  \right\}~.\label{Eq:OneLoopOpe_G}
\end{align}
To derive the above expressions, we use the equation of motion in the leading order of the expansion with respect to $1/M_4^2$, namely that of the SM.
In Eqs.~\eqref{Eq:OneLoopOpe_B}-\eqref{Eq:OneLoopOpe_G}, $F_B^{\mu\nu}, W^{I\mu\nu}$, and $G^{a\mu\nu}$ are the field strength of $\mathrm{U(1)_Y}$, SU(2), and $\mathrm{SU(3)_c}$ respectively.
The matching scale $\mu$ is typically taken to be VLQ mass scale $M_4$.
The lepton doublet is denoted by $l_L$.
The effective Lagrangians $\mathcal{L}_\mathrm{Eff}^B$, $\mathcal{L}_\mathrm{Eff}^W$ and $\mathcal{L}_\mathrm{Eff}^G$ contain the effective operators which contribute to the decay of $q_L^i\rightarrow q_L^jB$, $W^I$, and $G^a$, respectively. Since we use the equation of motions, the effective Lagrangians also contain the operators such as 4-Fermi operators which do not contribute to these processes.

Finally, the whole Lagrangian $\mathcal{L}_\mathrm{Eff}$ obtained by integrating out the VLQ fields is given as:
\begin{align}
  \mathcal{L}_\mathrm{Eff}
  &=
  \mathcal{L}_\mathrm{SM}
  +
  \mathcal{L}_\mathrm{Eff}^\mathrm{tree}
  +
  \mathcal{L}_\mathrm{Eff}^\mathrm{one-loop}~,\label{Eq:LagEff}\\
  \mathcal{L}_{\mathrm{SM}}&=
  \overline{q^i_L}i\Slash{D}_Lq_L^i+\overline{u_R^i}i\Slash{D}_R^uu_R^i+\overline{d_R^i}i\Slash{D}_R^dd_R^i
  -[y_d^{ij}\overline{q_L^i}\phi d_R^j
  +y_u^{ii}\overline{u_R^i}\tilde{\phi}q_L^i+\mathrm{h.c.}]~,\label{Eq:LagSM}
\end{align}
where $\mathcal{L}_\mathrm{Eff}^\mathrm{tree}$ is given in Eq.~\eqref{Eq:TreeOperator} or Eq.~\eqref{Eq:TreeOperator2} and
$\mathcal{L}_\mathrm{Eff}^\mathrm{one-loop}$ is given in Eqs.~\eqref{Eq:OneLoopOpe}-\eqref{Eq:OneLoopOpe_G}.
In Eq.~\eqref{Eq:LagEff}, the kinetic term of the SM quark doublet $q_L$ is
\begin{align}
  \mathcal{L}_\mathrm{K}^q &= \overline{q_L^j}\left\{\delta^{ji}+Z^{ji}(\mu)\right\}i\Slash{D}_{L}q_L^i~,\\[3pt]
  Z^{ji}(\mu) &= \frac{y_d^{j4}y_d^{i4*}}{16\pi^2}\left(\frac{1}{2}\ln\frac{\mu^2}{M_4^2}+\frac{3}{4}\right)~,
\end{align}
where the term $Z^{ji}(\mu)$ comes from the first term in Eq.~\eqref{Eq:OneLoopOpe}.
To rewrite the kinetic term into a canonical form, we perform the following rescaling of $q_L$
,
\begin{align}
  q_L'^k\equiv
  \left\{
  \delta^{ki} + \frac{1}{2}Z^{ki}(\mu)
  \right\}q_L^i~.\label{Eq.Rescale}
\end{align}
Then the kinetic term of the quark doublet becomes
\begin{align}
  \mathcal{L}_\mathrm{K}^q &= \overline{q_L'^k}i\Slash{D}_{L}q_L'^k~.
\end{align}
In terms of the rescaled fields introduced in Eq.~\eqref{Eq.Rescale}, the Yukawa interactions in Eq.~\eqref{Eq:LagSM} are changed:
\begin{align}
  y_u^{ii}\overline{q_L^i}\tilde{\phi}u_R^i
  &=
  \left\{
  \delta^{ki} - \frac{1}{2}Z^{ki}(\mu)
  \right\}y_u^{ii}\overline{q_L'^k}\tilde{\phi}u_R^i
  \equiv
  Y_u^{ki}\overline{q_L'^k}\tilde{\phi}u_R^i~,\label{Eq:Yu}\\
  y_d^{ji}\overline{q_L^j}\phi d_R^i
  &=
  \left\{
  \delta^{kj} - \frac{1}{2}Z^{kj}(\mu)
  \right\}
  y_d^{ji}\overline{q_L'^k}\phi d_R^i
  \equiv
  Y_d^{ki}\overline{q_L'^k}\phi d_R^i~,\label{Eq:Yd}
\end{align}
where we redefine the SM Yukawa coupling as
\begin{align}
  Y_u^{ki} \equiv
  \left\{
  \delta^{ki} - \frac{1}{2}Z^{ki}(\mu)
  \right\}y_u^{ii}~,
  \quad
  Y_d^{ki} \equiv
  \left\{
  \delta^{kj} - \frac{1}{2}Z^{kj}(\mu)
  \right\}y_d^{ji}~.
\end{align}
The rescaling of the field Eq.~\eqref{Eq.Rescale} is absorbed into the Yukawa couplings.
After the diagonalization of the mass matrices based on these couplings, it contributes to the CKM matrix as one-loop corrections. Since we only consider the charged current interaction in one-loop diagrams in the next section, these corrections lead to two-loop order effects and they are neglected.

The tree level effective operator in Eq.~\eqref{Eq:TreeOperator} is also changed by the rescaling in Eq.~\eqref{Eq.Rescale}:
\begin{align}
  i\frac{y_d^{j4}y_d^{i4*}}{M_4^2}
  \left(\overline{q_L^j}\phi\right)\Slash{D}_R^d\left(\phi^\dagger q_L^i\right)
  &=
  \left\{
  \delta^{kj} - \frac{1}{2}Z^{kj}(\mu)
  \right\}
  i\frac{y_d^{j4}y_d^{i4*}}{M_4^2}
  \left\{
  \delta^{il} - \frac{1}{2}Z^{il}(\mu)
  \right\}
  \left(\overline{q_L'^k}\phi\right)\Slash{D}_R^d\left(\phi^\dagger q_L'^l\right)\nonumber\\
  &\equiv
  i\frac{Y_d^{k4}Y_d^{l4*}}{M_4^2}
  \left(\overline{q_L'^k}\phi\right)\Slash{D}_R^d\left(\phi^\dagger q_L'^l\right)~,\label{Eq:Y4}
\end{align}
where we redefine the Yukawa coupling between the SM quarks and the VLQ as
\begin{align}
  Y_d^{k4}
  \equiv
  \left\{
  \delta^{kj} - \frac{1}{2}Z^{kj}(\mu)
  \right\}y_d^{j4}~.\label{Eq:Y42}
\end{align}
As we will see in the next subsection, this redefinition of the Yukawa coupling in Eq.~\eqref{Eq:Y42} adds $\mathcal{O}(\frac{1}{16\pi^2M_4^2})$ corrections to the CKM matrix and the FCNC coupling. In the next section we will take into account only leading order contributions in $1/M_4^2$, and these corrections are neglected.

For the one-loop effective Lagrangian, the rescaling in Eq.~\eqref{Eq.Rescale} leads to two-loop order corrections and we can simply take $q_L'\simeq q_L$ in the one-loop effective Lagrangian.

\subsection{Electroweak symmetry breaking}
In this subsection, we derive the Lagrangian for the broken phase of the SM gauge symmetry.
We substitute the following forms for the Higgs doublet in the Lagrangian Eqs.~\eqref{Eq:TreeOperator2}-\eqref{Eq:OneLoopOpe_G}:
\begin{align}
  \phi &=
  \begin{pmatrix}
    \chi^+\\
    (v+h+i\chi_0)/\sqrt{2}
  \end{pmatrix}~,\label{Eq:Higgsdoublet}\\[3pt]
  \tilde{\phi} &=
  \begin{pmatrix}
    (v+h-i\chi_0)/\sqrt{2}\label{Eq:Higgsdoublettilde}\\
      -\chi^-
  \end{pmatrix}~.
\end{align}
Here we do not take into account the running effect from the VLQ mass scale to the EW scale for the coefficients in the effective interactions Eqs.~\eqref{Eq:TreeOperator2}-\eqref{Eq:OneLoopOpe_G}.
For the effective Lagrangian $\mathcal{L}_\mathrm{Eff}^\mathrm{tree}$ in Eq.~\eqref{Eq:TreeOperator2}, we obtain
\begin{align}
  \mathcal{L}_\mathrm{Eff}^\mathrm{tree}
  &=
  \frac{v^2}{4M_4^2}
  \left[h_d^{ji}
  m_{d}^{ik}\overline{d_L^j}d_R^k + \mathrm{h.c.}
  \right]
  +\frac{g}{2M_W}\cdot\frac{3v^2}{4M_4^2}
  \left[h_d^{ji}
  m_d^{ik}\overline{d_L^j}d_R^k(h+i\chi_0)+\mathrm{h.c.}
  \right]\nonumber\\[3pt]
  &\phantom{=}
  -\frac{g}{\sqrt{2}M_W}\cdot\frac{v^2}{4M_4^2}
  \left[h_d^{ji}
  m_u^{ii}\overline{d_L^j}u_R^i\chi^- + \mathrm{h.c.}
  \right]
  +\frac{g}{\sqrt{2}M_W}\cdot\frac{v^2}{2M_4^2}
  \left[h_d^{ji}
  m_d^{ik}\overline{u_L^j}d_R^k\chi^+ +\mathrm{h.c.}
  \right]\nonumber\\[3pt]
  &\phantom{=}
  +\frac{g}{\sqrt{2}}\cdot\frac{v^2}{4M_4^2}
  \left[h_d^{ji}
  \overline{u_L^j}\gamma^\mu d_L^iW_\mu^+ +\mathrm{h.c.}
  \right]
  -\frac{g}{2c_w}\cdot\frac{v^2}{2M_4^2}
  h_d^{ji}
  \overline{d_L^j}\gamma^\mu d_L^iZ_\mu
  +\cdots~,\label{Eq:TreeOperatorBP}
\end{align}
where the ellipsis represents the terms including more than four fields, $h_d^{ji}$ represents $y_d^{j4}y_d^{i4*}$ and
$c_w$ $(s_w)$ denotes cosine (sine) of the weak mixing angle $\theta_w$.
The mass matrices of the up-type and down-type quarks which correspond to $\mathcal{L}_\mathrm{SM}$ in Eq.~\eqref{Eq:LagSM} are denoted by $m_{u,d}\equiv vy_{u,d}/\sqrt{2}$.
Adding the tree level effective Lagrangian Eq.~\eqref{Eq:TreeOperatorBP} to the SM Lagrangian $\mathcal{L}_\mathrm{SM}$ in Eq.~\eqref{Eq:LagSM}, the mass matrix of the down-type quarks
changes into $\left(\delta^{ji}-\frac{v^2}{4M_4^2}h_d^{ji}\right)m_d^{ik}$.
We diagonalize this mass matrix.
At first, we introduce $3\times 3$ unitary matrices $K_L$ and $K_R$. These unitary matrices diagonalizes the matrix $m_d$:
\begin{align}
  \left\{
  \begin{array}{ll}
    d_L^i = K_L^{im}d_L'^m\\[3pt]
    d_R^i = K_R^{im}d_R'^m
  \end{array}
  \right.
  \quad
  \rightarrow
  \quad
  (K_L^\dagger m_d K_R)^{mn} \equiv m_d'^{m}\delta^{mn}
  ~,\label{Eq:UnitaryTrans1}
\end{align}
where the prime indicates the mass basis of the SM.
In this mass basis, the mass matrix of the down-type quarks changes into
\begin{align}
  K_L^{\dagger mj}\left(\delta^{ji}-\frac{v^2}{4M_4^2}h_d^{ji}\right)m_d^{ik}K_R^{kn}
  =
  \left(\delta^{mn}-\frac{v^2}{4M_4^2}h_d'^{mn}\right)m_d'^{n}~,\label{Eq:MassMatrixPrime}
\end{align}
where
\begin{align}
  (K_L^\dagger h_d K_L)^{mn} = K_L^{\dagger mj}y_d^{j4}y_d^{i4*}K_L^{in}
  \equiv y_d'^{m4}y_d'^{n4*} \equiv h_d'^{mn}~\label{Eq:hdp}.
\end{align}
The mass matrix in Eq.~\eqref{Eq:MassMatrixPrime} is not diagonal.
In order to diagonalize the mass matrix including contribution of $\mathcal{O}(v^2/M_4^2)$, we introduce unitary matrices $V_L$ and $V_R$,
\begin{align}
  \left\{
  \begin{array}{ll}
    d_L'^m = V_L^{mp}d_L''^p\\[3pt]
    d_R'^m = V_R^{mp}d_R''^p
  \end{array}
  \right.
  \quad
  \rightarrow
  \quad
  V_L^{\dagger pm}
  \left(\delta^{mn}-\frac{v^2}{4M_4^2}h_d'^{mn}\right)m_d'^{n}V_R^{nq}
  \equiv
  m_d''^{p}\delta^{pq}~,\label{Eq:UnitaryTrans2}
\end{align}
where the double prime denotes the mass basis of the model with VLQ.
The physical masses of the down-type quarks are denoted by $m_d''^p = (m_d'',m_s'',m_b'')$.
The mixing angles of these unitary matrices are the order of $\mathcal{O}(v^2/M_4^2)$.
Hereafter we omit the double prime on the quark fields of the mass basis and $h_d$ denotes the $h_d'$ in the right-hand side of Eq.~\eqref{Eq:hdp}.
Finally, we obtain the following Lagrangian after the transformation in Eqs.~\eqref{Eq:UnitaryTrans1} and \eqref{Eq:UnitaryTrans2},
\begin{align}
  \mathcal{L}_\mathrm{SM} + \mathcal{L}_\mathrm{Eff}^\mathrm{tree}
  =
  \mathcal{L}_0 + \mathcal{L}_{A} + \mathcal{L}_{W} + \mathcal{L}_{Z} + \mathcal{L}_{\chi^\pm} + \mathcal{L}_{h} + \mathcal{L}_{\chi_0} + \cdots ~,\label{Eq:LVLQ}
\end{align}
where the ellipsis represents the terms which contain more than four fields.
Each part of the Lagrangians is given below:
\begin{align}
  \mathcal{L}_0
  &=
  \overline{u^i}i\Slash{\partial}u^i + \overline{d^p}i\Slash{\partial}d^p
  -\left[
  m_u^{i}\overline{u^i}u^i + m_d^p\overline{d^p}d^p
  \right]~,\label{Eq:L0}\\[3pt]
  \mathcal{L}_{A}
  &=
  -e\left[Q_u\overline{u^i}\gamma^\mu u^i
  + Q_d\overline{d^p}\gamma^\mu d^p\right]A_\mu ~,\label{Eq:LA}\\[3pt]
  \mathcal{L}_{W}
  &=
  -\frac{g}{\sqrt{2}}
  \overline{u^i} \gamma^\mu V_\mathrm{CKM}^{iq}L d^q W_\mu^+ +\mathrm{h.c.}~,\label{Eq:LW}\\[3pt]
  \mathcal{L}_{Z}
  &=
  -\frac{g}{c_w}\left[\overline{u^i}\gamma^\mu
  \left(\frac{1}{2}L-Q_us_w^2\right) u^i
  -\overline{d^p}\gamma^\mu
  \left(\frac{1}{2}Z_\mathrm{NC}^{pq}L+Q_ds_w^2\delta^{pq}\right) d^q
  \right]Z_\mu ~,\label{Eq:LZ}\\[3pt]
  \mathcal{L}_{\chi^\pm}
  &=
  \frac{g}{\sqrt{2}M_W}
  \overline{u^i}V_\mathrm{CKM}^{ip}
  \left(
  m_u^{i}L-m_d^{p}R
  \right)d^p\chi^+ + \mathrm{h.c.}~,\label{Eq:Lchipm}\\[3pt]
  \mathcal{L}_{h}
  &=
  -\frac{g}{2M_W}\overline{d^p}Z_\mathrm{NC}^{pq}\left(m_d^{q}R+m_d^{p}L\right)d^qh~,\label{Eq:Lh}\\[3pt]
  \mathcal{L}_{\chi_0}
  &=
  -\frac{ig}{2M_W}
  \overline{d^p}Z_\mathrm{NC}^{pq}\left(m_d^{q}R-m_d^{p}L\right)d^q\chi_0 ~.\label{Eq:Lchi0}
\end{align}
In Eqs.~\eqref{Eq:L0}-\eqref{Eq:Lchi0}, $L$ and $R$ denote the chiral projection operators, $L\equiv\frac{1-\gamma_5}{2}, R\equiv\frac{1+\gamma_5}{2}$. The electromagnetic charge of up-type and down-type quarks are denoted by $Q_u$ and $Q_d$ respectively. The $3\times 3$ CKM matrix $V_\mathrm{CKM}$ is defined as,
\begin{align}
  V_\mathrm{CKM} \equiv K_L\left(1-\frac{v^2}{4M_4^2}h_d\right)V_L~.\label{Eq:CKMMatrix}
\end{align}
The FCNCs arise from the $3\times 3$ non-diagonal matrix $Z_\mathrm{NC}$ in the $Z$, $h$ and $\chi_0$ interactions in Eqs.~\eqref{Eq:LZ},~\eqref{Eq:Lh} and \eqref{Eq:Lchi0}. The matrix $Z_{NC}$ in the neutral currents is defined as follows:
\begin{align}
  Z_\mathrm{NC}
  \equiv
  V_L^\dagger
  \left(
  1-\frac{v^2}{2M_4^2}h_d
  \right)V_L
  \simeq
  1-\frac{v^2}{2M_4^2}h_d + \mathcal{O}(v^4/M_4^{4})~.\label{Eq:ZNC}
\end{align}
Using Eqs.~\eqref{Eq:CKMMatrix} and \eqref{Eq:ZNC}, we obtain the relation between the CKM matrix $V_\mathrm{CKM}$ and the matrix $Z_\mathrm{NC}$ up to $\mathcal{O}(v^2/M_4^2)$,
\begin{align}
  \sum_{i=u,c,t}V_\mathrm{CKM}^{ip*}V_\mathrm{CKM}^{iq}
  =
  Z_\mathrm{NC}^{pq}~.\label{Eq:UnitarityRelation}
\end{align}
Equation~\eqref{Eq:UnitarityRelation} shows that the unitarity of the CKM matrix for the three generations does not hold due to the deviation from the unit matrix of the matrix $Z_\mathrm{NC}$ in Eq.~\eqref{Eq:ZNC}.
Taking the limit of $M_4 \rightarrow \infty$, the unitarity relation is restored.

Next we rewrite the one-loop level effective Lagrangian Eqs.~\eqref{Eq:OneLoopOpe}-\eqref{Eq:OneLoopOpe_G} in terms of the mass basis defined in Eq.~\eqref{Eq:UnitaryTrans2}. Below we write the part of the dipole operators and omit the other parts of the effective Lagrangian:
\begin{align}
  \phantom{=}\mathcal{L}_\mathrm{Eff}^B + \mathcal{L}_\mathrm{Eff}^W
  &=
  \frac{g}{16\pi^2c_w}\cdot\frac{G_F}{6\sqrt{2}}
  \left(
  1-\frac{7}{2}Q_us_w^2
  \right)
  \overline{u_L^j}V_\mathrm{CKM}^{jp}
  \left(
  \delta^{pq}-Z_{\mathrm{NC}}^{pq}
  \right)V_\mathrm{CKM}^{iq*}m_u^i
  \sigma_{\mu\nu}u_R^iZ^{\mu\nu}\nonumber\\[3pt]
  &\phantom{=}+
  \frac{g}{16\pi^2c_w}\cdot\frac{G_F}{6\sqrt{2}}
  \left(
  -1+Q_ds_w^2
  \right)
  \overline{d_L^p}\left(\delta^{pq}-Z_\mathrm{NC}^{pq}\right)m_d^{q}\sigma_{\mu\nu}d_R^qZ^{\mu\nu}\nonumber\\[3pt]
  &\phantom{=}+
  \frac{e}{16\pi^2}\cdot\frac{7G_F}{12\sqrt{2}}
  Q_u\overline{u_L^j}V_\mathrm{CKM}^{jp}
  \left(
  \delta^{pq}-Z_\mathrm{NC}^{pq}
  \right)V_\mathrm{CKM}^{iq*}m_u^i\sigma_{\mu\nu}u_R^i
  F_A^{\mu\nu}\nonumber\\[3pt]
  &\phantom{=}-
  \frac{e}{16\pi^2}\cdot\frac{G_F}{6\sqrt{2}}Q_d\overline{d_L^p}\left(\delta^{pq}-Z_\mathrm{NC}^{pq}
  \right)m_d^q\sigma_{\mu\nu}d_R^q
  F_A^{\mu\nu}\nonumber\\[3pt]
  &\phantom{=}+
  \frac{g}{16\sqrt{2}\pi^2}\cdot\frac{G_F}{3\sqrt{2}}
  \overline{d_L^p}\left(\delta^{pq}-Z_\mathrm{NC}^{pq}\right)V_\mathrm{CKM}^{iq*}m_u^i\sigma_{\mu\nu}u_R^iW^{-\mu\nu}\nonumber\\[3pt]
  &\phantom{=}+
  \frac{g}{16\sqrt{2}\pi^2}\cdot\frac{G_F}{3\sqrt{2}}
  \overline{u_L^j}V_\mathrm{CKM}^{jp}
  \left(
  \delta^{pq}-Z_\mathrm{NC}^{pq}
  \right)m_d^q\sigma_{\mu\nu}d_R^qW^{+\mu\nu} +\mathrm{h.c.}~,\label{Eq:DipoleEW}
\end{align}
\begin{align}
  \mathcal{L}_\mathrm{Eff}^G
  &=
  -\frac{g_s}{16\pi^2}\cdot\frac{G_F}{6\sqrt{2}}
  \overline{u_L^j}V_\mathrm{CKM}^{jp}
  \left(\delta^{pq}-Z_\mathrm{NC}^{pq}\right)
  V_\mathrm{CKM}^{iq*}m_u^i\frac{\lambda^a}{2}\sigma_{\mu\nu}u_R^i
  G^{a\mu\nu}\nonumber\\[3pt]
  &\phantom{=}
  -\frac{g_s}{16\pi^2}\cdot\frac{G_F}{6\sqrt{2}}
  \overline{d_L^p}
  \left(\delta^{pq}-Z_\mathrm{NC}^{pq}\right)
  m_d^{q}\frac{\lambda^a}{2}\sigma_{\mu\nu}d_R^q
  G^{a\mu\nu} +\mathrm{h.c.}~,\label{Eq:DipoleGluon}
\end{align}
where the field strength $Z^{\mu\nu},F_A^{\mu\nu}$ and $W^{\pm\mu\nu}$ are defined as,
\begin{align}
  Z^{\mu\nu}
  =
  \partial^\mu Z^\nu - \partial^\nu Z^\mu~,\quad
  F_A^{\mu\nu}
  =
  \partial^\mu A^\nu - \partial^\nu A^\mu~,\quad
  W^{\pm\mu\nu}
  =
  \partial^\mu W^{\pm\nu} - \partial^\nu W^{\pm\mu}~,
\end{align}
respectively.
Note that the coefficient of the photon dipole operator with the down-type quarks is consistent with the case of the full theory calculation up to $\mathcal{O}(M_4^{-2})$ \cite{Handoko:1994xw}.

\section{Effective Lagrangian for $\Delta B=1,2$ and $b\rightarrow s\gamma^{(*)}$ processes}
\label{sec:3}

In order to analyze the $B$ meson system, we derive the effective Lagrangian for $\Delta B=1,2$ and $b\rightarrow s\gamma^{(*)}$ processes in the model with VLQ.
Here we focus on contributions derived from the effective Lagrangian in Eq.~\eqref{Eq:LVLQ}.
There are three sources to the effective Lagrangian.
The first contribution is the same as the case of the SM.
The second contribution corresponds to the diagrams which include the FCNC couplings.
The third contribution comes from the violation of the CKM unitarity.
In the following computations, we use the 't~Hooft-Feynman gauge.
\subsection{$\Delta B = 1$ process}
\label{subsec:3.1}

\begin{figure}[tbp]
  \centering
  \includegraphics[width=16.5cm]{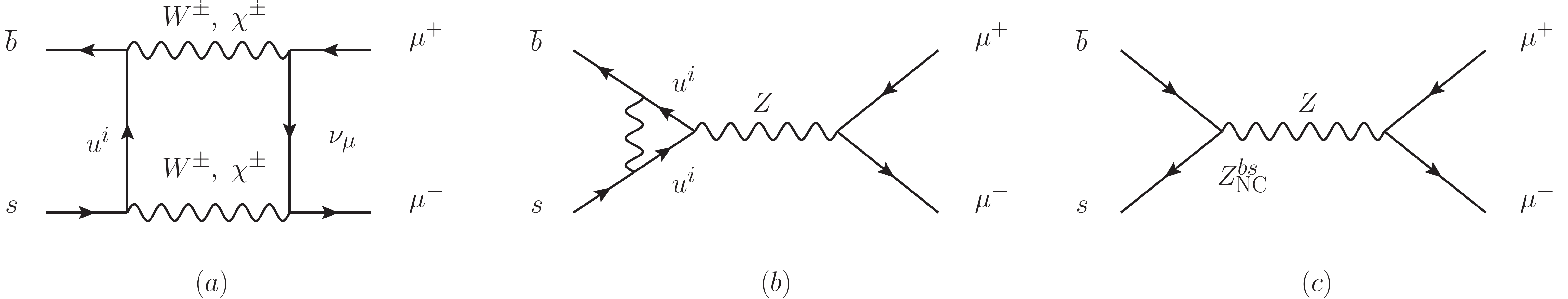}
  \caption{The Feynman diagrams which contribute to the $\overline{b}s\rightarrow \mu^+\mu^-$ process. The diagrams $(a)$ and $(b)$ are the same contributions as the SM. The diagram $(c)$ is the new contribution in the model with VLQ.}
  \label{Fig:bsll}
\end{figure}

At first we consider the $\Delta B = 1$ process to calculate the branching ratio of the $B_s\rightarrow \mu^+\mu^-$ process in the next section. The diagrams which contribute to the effective Lagrangian up to $\mathcal{O}(Z_\mathrm{NC})$ are shown in Fig.~\ref{Fig:bsll}.
The $\bar{b}s\rightarrow \mu^+\mu^-$ process occurs at tree level in the model with VLQ since there is the tree level $Z$ FCNC among the down-type quarks.
In one-loop level, the contribution comes from the Feynman diagrams in Fig.~\ref{Fig:bsll}$(a)$ and Fig.~\ref{Fig:bsll}$(b)$ which are also present in the SM.
However these amplitudes include the additional contributions due to the violation of the CKM unitarity.
Since these contributions are suppressed by the loop factor $e^2/(16\pi^2)$ compared with the contribution of the tree diagram in Fig.~\ref{Fig:bsll}$(c)$, we neglect the contribution from the violation of the CKM unitarity in the computation of the $\bar{b}s\rightarrow \mu^+\mu^-$ process.
Then the effective Lagrangian for the $\overline{b}s\rightarrow \mu^+\mu^-$ is given as follows:
\begin{align}
  \mathcal{L}_\mathrm{Eff}(\overline{b}s\rightarrow\mu^+\mu^-)
  =
  \frac{\sqrt{2}G_F\alpha_{em}}{\pi s_w^2}
  \lambda_{bs}^tY_0(x_t)
  \left\{
  1-\frac{\pi s_w^2}{\alpha_{em}Y_0(x_t)}\cdot\frac{Z_\mathrm{NC}^{bs}}{\lambda_{bs}^t}
  \right\}
  \left[
  \overline{b_L}\gamma^\mu s_L
  \right]
  \left[\overline{\mu_L}\gamma_\mu\mu_L
  \right]~,\label{Eq:Leffbsll}
\end{align}
where $\lambda_{bs}^t\equiv V_\mathrm{CKM}^{tb*}V_\mathrm{CKM}^{ts}$, and $\alpha_{em} = e^2/(4\pi)$ denote the fine structure constant of the electromagnetic interaction. The Inami-Lim function $Y_0(x_t)$ is \cite{Inami:1980fz,Buchalla:1990qz},
\begin{align}
  Y_0(x_i)
  =
  \frac{1}{8}x_i - \frac{3}{8}\frac{x_i}{x_i-1} + \frac{3}{8}\frac{x_i^2}{(x_i-1)^2}\ln x_i~,\label{Eq:ILfunctionY}
\end{align}
where $x_i\equiv (m_u^i/M_W)^2$. The first term in Eq.~\eqref{Eq:Leffbsll} comes from the diagrams in Fig.~\ref{Fig:bsll}$(a)$ and Fig.~\ref{Fig:bsll}$(b)$ with the CKM unitarity relation for the SM $(\sum_{i=u,c,t}\lambda_{bs}^i = 0)$, that is the SM contribution. The second term of Eq.~\eqref{Eq:Leffbsll} comes from the diagram in Fig.~\ref{Fig:bsll}$(c)$, so this term is the new contribution in the model with VLQ.
\subsection{$\Delta B = 2$ process}
\label{subsec:3.2}

\begin{figure}[tbp]
  \centering
  \includegraphics[width=16.5cm]{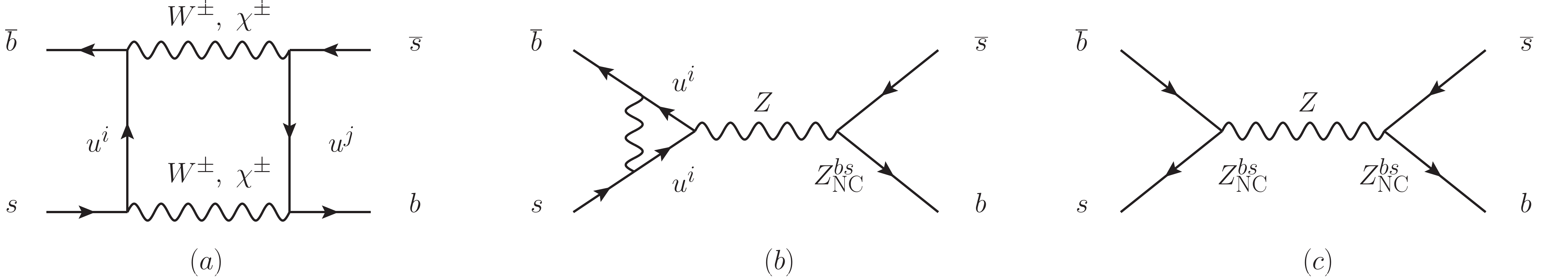}
  \caption{The Feynman diagrams which contribute to the $\Delta B = 2$ process in the model with VLQ. The diagram $(a)$ is same in the case of the SM, but the additional contribution arises from the diagram $(a)$ by the violation of the CKM unitarity.}
  \label{Fig:Bsmixing}
\end{figure}

Next we
show
the effective Lagrangian for $\Delta B = 2$ process in order to compute the mass difference of $B_s$ meson in the later section.
In Refs.~\cite{Branco:1992wr}-\cite{Barenboim:1997pf}, $\Delta B = 2$ process was computed up to $\mathcal{O}(Z_\mathrm{NC}^2)$ and $\mathcal{O}(Z_\mathrm{NC}\cdot\alpha_{em}/(4\pi))$.
The diagrams which contribute to the $\Delta B = 2$ process are given in Fig.~\ref{Fig:Bsmixing}.
From the diagrams in Fig.~\ref{Fig:Bsmixing}, we obtain the effective Lagrangian for $\overline{b}s\leftrightarrow\overline{s}b$ process as \cite{Branco:1992wr}-\cite{Barenboim:1997pf}:
\begin{align}
  \mathcal{L}_\mathrm{Eff}(\overline{b}s\leftrightarrow\overline{s}b)
  &=
  \frac{G_F}{\sqrt{2}}\frac{\alpha_{em}}{4\pi s_w^2}\left(\lambda_{bs}^t\right)^2S_0(x_t)\nonumber\\[3pt]
  &\phantom{=}
  \times
  \left\{
  1+\frac{8Y_0(x_t)}{S_0(x_t)}\cdot\frac{Z_\mathrm{NC}^{bs}}{\lambda_{bs}^t}
  -\frac{4\pi s_w^2}{\alpha_{em}S_0(x_t)}\left(\frac{Z_\mathrm{NC}^{bs}}{\lambda_{bs}^t}\right)^2
  \right\}
  \left[\overline{b_L}\gamma^\mu s_L\right]\left[\overline{b_L}\gamma_\mu s_L\right]~,\label{Eq:LeffbsMixing}
\end{align}
where
\begin{align}
  S_0(x_i)
  =
  -\frac{3}{2}\left(\frac{x_i}{x_i-1}\right)^3\ln x_i
  -
  x_i\left\{
  \frac{1}{4}-\frac{9}{4}\frac{1}{x_i-1}-\frac{3}{2}\frac{1}{(x_i-1)^2}
  \right\}~,\label{IfunctionS}
\end{align}
is the Inami-Lim function \cite{Inami:1980fz}.
Also $Y_0(x_t)$ is given in Eq.~\eqref{Eq:ILfunctionY}.
The first term in Eq.~\eqref{Eq:LeffbsMixing} comes from the diagram in Fig.~\ref{Fig:Bsmixing}$(a)$ with the CKM unitarity relation.
The second term in Eq.~\eqref{Eq:LeffbsMixing} is obtained from the violation of the CKM unitarity in the diagram in Fig.~\ref{Fig:Bsmixing}$(a)$ in addition to the contribution from the diagram in Fig.~\ref{Fig:Bsmixing}$(b)$.
The CKM unitarity relation is used for the one-loop $Z$ FCNC vertex in Fig.~\ref{Fig:Bsmixing}$(b)$ since $\mathcal{O}(Z_\mathrm{NC}^2\cdot\alpha_{em}/(4\pi))$ contribution is neglected.
The third term comes from the diagram in Fig.~\ref{Fig:Bsmixing}$(c)$.
Note that the effective Lagrangian in Eq.~\eqref{Eq:LeffbsMixing} contains only the Inami-Lim functions which are gauge-parameter independent \cite{Barenboim:1997pf}.
\subsection{$b\rightarrow s\gamma^{(*)}$ process}
\label{subsec:3.3}

\begin{figure}[tbp]
  \centering
  \includegraphics[width=16.5cm]{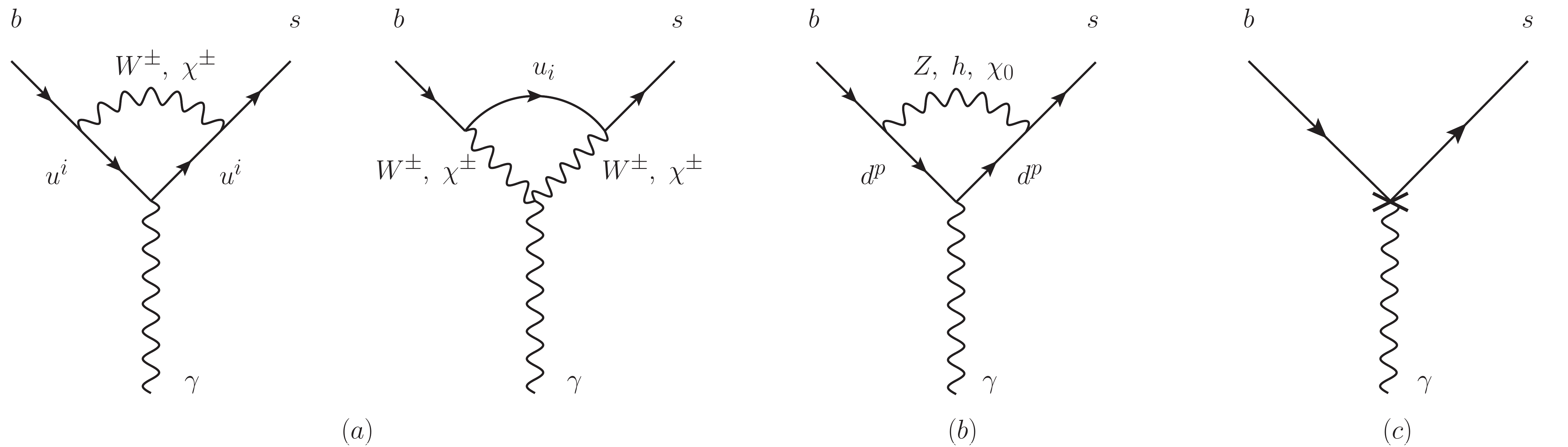}
  \caption{The Feynman diagrams which contribute to the $b\rightarrow s\gamma^{(*)}$ process. The diagrams $(a)$ are the same contributions as the SM. The additional contribution arises from the diagrams $(a)$ by the violation of the CKM unitarity. The diagram $(b)$ is the new contribution in the model with VLQ. The diagram $(c)$ is the counterterm determined by the quark-self energy and $Z$-photon, $\chi_0$-photon mixing diagrams in Fig.~\ref{Fig:ZAMixing}.}
  \label{Fig:bsgamma}
\end{figure}

Finally we derive the effective Lagrangian for $b\rightarrow s\gamma^{(*)}$ process to evaluate the $\bar{B}$ meson radiative decay $\bar{B}\rightarrow X_s\gamma$.
In addition to the contribution from the effective Lagrangian in Eq.~\eqref{Eq:DipoleEW}, the diagrams in Fig.~\ref{Fig:bsgamma} are also contribute to the $b\rightarrow s\gamma^{(*)}$ process.
The effective Lagrangian for $b\rightarrow s\gamma$ process was calculated in terms of the full theory \cite{Handoko:1994xw,Chang:1998sk} while the effective Lagrangian for $b\rightarrow s\gamma^{*}$ process was not calculated.
Here we will give the effective Lagrangian for both $b\rightarrow s\gamma$ and  $b\rightarrow s\gamma^{*}$ in terms of the effective theory.

\begin{figure}[tbp]
  \centering
  \includegraphics[width=7.0cm]{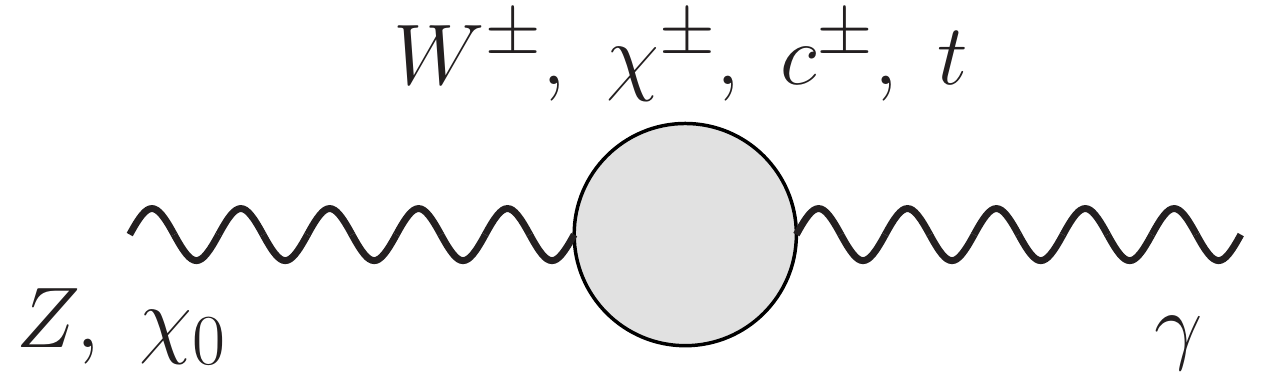}
  \caption{The diagram where photon mixes with $Z$ or $\chi_0$ at one-loop level. $c^\pm$ denotes the Faddeev-Popov ghost.}
  \label{Fig:ZAMixing}
\end{figure}

In the model with VLQ, there are no FCNC by the quark-quark-photon interaction at tree level. Therefore the leading order contributions which contain the FCNC couplings come from the violation of the CKM unitarity in the diagrams in Fig.~\ref{Fig:bsgamma}$(a)$ and the one-loop diagram in Fig.~\ref{Fig:bsgamma}$(b)$ \cite{Handoko:1994xw,Chang:1998sk}.
In order to obtain the effective Lagrangian for $b\rightarrow s\gamma^{(*)}$ process, we compute the amplitudes of the diagrams in Fig.~\ref{Fig:bsgamma}. We introduce several counterterms when we renormalize amplitudes for Figs.~\ref{Fig:bsgamma}$(a)$ and \ref{Fig:bsgamma}$(b)$.
As mentioned in Ref.~\cite{Handoko:1994xw}, the counterterms of the renormalization for the quark fields remove the divergence of the diagrams in Fig.~\ref{Fig:bsgamma}$(a)$ with the CKM unitarity and that of the diagram in Fig.~\ref{Fig:bsgamma}$(b)$.
However these counterterms cannot remove all the divergence arising from these diagrams.
There still remains the divergence which comes from the violation of the CKM unitarity in Fig.~\ref{Fig:bsgamma}$(a)$.
Therefore we have to introduce another counterterm. We consider the renormalization for the neutral gauge bosons $Z$ and $A$ \cite{Aoki:1982ed}. The bare fields $Z_0^\mu$ and $A_0^\mu$ are related to the renormalized fields as follows:
\begin{align}
  \begin{pmatrix}
    Z_0^\mu\\
    A_0^\mu
  \end{pmatrix}
  =
  \begin{pmatrix}
    \sqrt{Z_\mathrm{ZZ}} & \sqrt{Z_\mathrm{ZA}}\\
    \sqrt{Z_\mathrm{AZ}} & \sqrt{Z_\mathrm{AA}}
  \end{pmatrix}
  \begin{pmatrix}
    Z^\mu\\
    A^\mu
  \end{pmatrix}~,\label{Eq:RenormalizationZA}
\end{align}
where $\sqrt{Z_{ij}},(i,j=Z,A)$ are the renormalization constants. The divergence coming from the violation of the CKM unitarity in the diagrams in Fig.~\ref{Fig:bsgamma}$(a)$ is exactly cancelled by the counterterm given as
\begin{align}
  Z_\mathrm{NC}^{sb}\bar{s}\gamma_\mu LbZ_0^\mu
  \quad
  \rightarrow
  \quad
  \sqrt{Z_\mathrm{ZA}}\cdot Z_\mathrm{NC}^{sb}\bar{s}\gamma_\mu LbA^\mu~.
\end{align}
The renormalization constant $\sqrt{Z_\mathrm{ZA}}$ and $\sqrt{Z_\mathrm{AZ}}$ are determined by the diagrams in Fig.~\ref{Fig:ZAMixing} where $Z$ (or $\chi_0$) and photon mix at one-loop level.
The finite part of the transition amplitude of the diagram in Fig.~\ref{Fig:ZAMixing} contribute to the effective Lagrangian for the $b\rightarrow s\gamma^{*}$ process. Finally, we can obtain the effective Lagrangian $\mathcal{L}_\mathrm{Eff}(b\rightarrow s\gamma)$ for the on-shell photon and the effective Lagrangian $\mathcal{L}_\mathrm{Eff}(b\rightarrow s\gamma^{*})$ which vanishes for the on-shell photon as follows:
\begin{align}
  \mathcal{L}_\mathrm{Eff}(b\rightarrow s\gamma)
  &=
  \mathcal{L}_\mathrm{Eff}^\mathrm{CC}(b\rightarrow s\gamma)
  +
  \mathcal{L}_\mathrm{Eff}^\mathrm{uv}(b\rightarrow s\gamma)
  +
  \mathcal{L}_\mathrm{Eff}^\mathrm{NC}(b\rightarrow s\gamma)~,\label{Eq:Leffbsgamma}\\[3pt]
  \mathcal{L}_\mathrm{Eff}(b\rightarrow s\gamma^*)
  &=
  \mathcal{L}_\mathrm{Eff}^\mathrm{CC}(b\rightarrow s\gamma^*)
  +
  \mathcal{L}_\mathrm{Eff}^\mathrm{uv}(b\rightarrow s\gamma^*)
  +
  \mathcal{L}_\mathrm{Eff}^\mathrm{NC}(b\rightarrow s\gamma^*)
  +
  \mathcal{L}_\mathrm{Eff}^\mathrm{Mix}(b\rightarrow s\gamma^*)
  ~,\label{Eq:Leffbsgammaast}
\end{align}
where the indices ``CC''denote the contributions from the diagrams in Fig.~\ref{Fig:bsgamma}$(a)$ with the CKM unitarity relation, namely the SM contributions. Also indices ``uv'' and ``NC'' imply the contributions from the violation of the CKM unitarity and the diagram in Fig.~\ref{Fig:bsgamma}$(b)$ which include the neutral current respectively.
The index ``Mix'' indicates the contributions from the $Z$-photon and $\chi_0$-photon mixing diagrams. Concretely these effective Lagrangian are obtained as:
\begin{align}
  \mathcal{L}_\mathrm{Eff}^\mathrm{CC}(b\rightarrow s\gamma)
  &=
  -\frac{G_Fe}{8\sqrt{2}\pi^2}
  \sum_{i=c,t}\lambda_{sb}^i
  \left\{
  Q_uF_u(x_i) + F_W(x_i)
  \right\}
  \overline{s}\sigma_{\mu\nu}\left(m_bR+m_sL\right)bF_A^{\mu\nu}~,
  \label{Eq:LeffbsgammaCC}\\[3pt]
  \mathcal{L}_\mathrm{Eff}^\mathrm{uv}(b\rightarrow s\gamma)
  &=
  \frac{G_Fe}{8\sqrt{2}\pi^2}Z_\mathrm{NC}^{sb}
  \left(\frac{2}{3}Q_u+\frac{5}{6}\right)
  \overline{s}\sigma_{\mu\nu}\left(m_bR+m_sL\right)bF_A^{\mu\nu}~,
  \label{Eq:Leffbsgammauv}\\[3pt]
  \mathcal{L}_\mathrm{Eff}^\mathrm{NC}(b\rightarrow s\gamma)
  &=
  \frac{G_Fe}{8\sqrt{2}\pi^2}Q_d
  \sum_{p=d,s,b}
  \Bigl\{
  Z_\mathrm{NC}^{sp}Z_\mathrm{NC}^{pb}F_\mathrm{ZZ}(r_p,w_p)\Bigr.
  \nonumber\\[3pt]
  &\phantom{=+}\Bigl.
  +Z_\mathrm{NC}^{sb}Q_ds_w^2(\delta^{sp}+\delta^{pb})F_\mathrm{Z}(r_p)
  \Bigr\}
  \overline{s}\sigma_{\mu\nu}\left(m_bR+m_sL\right)bF_A^{\mu\nu}\nonumber\\[3pt]
  &\phantom{=}
  -\frac{G_Fe}{4\sqrt{2}\pi^2}Q_d
  \sum_{p=s,b}
  Z_\mathrm{NC}^{sb}Q_ds_w^2F_\mathrm{Z}'(r_p)
  \overline{s}\sigma_{\mu\nu}\left(\delta^{pb}m_bR+\delta^{sp}m_sL\right)bF_A^{\mu\nu}~,\label{Eq:LeffbsgammaNC}
\end{align}
and
\begin{align}
  \mathcal{L}_\mathrm{Eff}^\mathrm{CC}(b\rightarrow s\gamma^*)
  &=
  -\frac{G_Fe}{8\sqrt{2}\pi^2}
  \sum_{i=c,t}\lambda_{sb}^i
  \left\{
  Q_uf_u(x_i)+f_W(x_i)
  \right\}\overline{s}\gamma_\nu Lb \partial_\mu F_A^{\mu\nu}~,\label{Eq:LeffbsgammaastCC}\\[3pt]
  \mathcal{L}_\mathrm{Eff}^\mathrm{uv}(b\rightarrow s\gamma^*)
  &=
  -\frac{G_Fe}{8\sqrt{2}\pi^2}
  Z_\mathrm{NC}^{sb}
  \left\{
  Q_u\left(-\frac{2}{9}+\frac{4}{3}\ln x_u\right)-\frac{16}{9}
  \right\}\overline{s}\gamma_\nu Lb \partial_\mu F_A^{\mu\nu}~,\\[3pt]
  \mathcal{L}_\mathrm{Eff}^\mathrm{NC}(b\rightarrow s\gamma^*)
  &=
  \frac{G_Fe}{8\sqrt{2}\pi^2}Q_d
  \sum_{p=d,s,b}
  \Bigl\{
  Z_\mathrm{NC}^{sp}Z_\mathrm{NC}^{pb}f_\mathrm{ZZ}(r_p,w_p)
  \nonumber\\[3pt]
  &\phantom{=+}
  +Z_\mathrm{NC}^{sb}Q_ds_w^2\left(\delta^{sp}+\delta^{pb}\right)f_\mathrm{Z}(r_p)
  \Bigr\}\overline{s}\gamma_\nu Lb \partial_\mu F_A^{\mu\nu}~,\label{Eq:LeffbsgammaastNC}\\[3pt]
  \mathcal{L}_\mathrm{Eff}^\mathrm{Mix}(b\rightarrow s\gamma^*)
  &=
  \frac{G_Fe}{8\sqrt{2}\pi^2}Z_\mathrm{NC}^{sb}
  \left\{
  \left(10c_w^2 + \frac{1}{3}\right)\ln\frac{\mu^2}{M_W^2} + \frac{4}{3}c_w^2
  \right.\nonumber\\[3pt]
  &\phantom{=+}\left.
  -2Q_u\left(1-4Q_us_w^2\right)\ln\frac{\mu^2}{m_t^2}
  \right\}\overline{s}\gamma_\nu Lb \partial_\mu F_A^{\mu\nu}~.
\end{align}
The Inami-Lim functions in Eqs.~\eqref{Eq:LeffbsgammaCC} and \eqref{Eq:LeffbsgammaastCC} are given as follows \cite{Inami:1980fz}:
\begin{align}
  F_u(x_i)
  &\equiv
  \frac{x_i(2+3x_i-6x_i^2+x_i^3+6x_i\ln x_i)}{4(x_i-1)^4}~,\\[3pt]
  F_W(x_i)
  &\equiv
  \frac{x_i(1-6x_i+3x_i^2+2x_i^3-6x_i^2\ln x_i)}{4(x_i-1)^4}~,\\[3pt]
  f_u(x_i)
  &\equiv
  -\frac{x_i\{18-29x_i+10x_i^2+x_i^3+(32-18x_i)\ln x_i\}}{6(x_i-1)^4}
  +\frac{4}{3(x_i-1)^4}\ln x_i-\frac{4}{3}\ln x_u~,\\[3pt]
  f_W(x_i)
  &\equiv
  \frac{x_i\{12-11x_i-8x_i^2+7x_i^3+2x_i(12-10x_i+x_i^2)\ln x_i\}}{6(x_i-1)^4}~,
\end{align}
where the subscripts ``$u$'' and ``$W$'' indicate the contributions which are proportional to the electromagnetic charge of the up-type quarks and the $W$ boson respectively. The functions $F_\mathrm{ZZ}$, $F_\mathrm{Z}$ and $F_\mathrm{Z}'$ in Eq.~\eqref{Eq:LeffbsgammaNC} are given as\footnote{The terms linear to $Z_\mathrm{NC}^{sb}$ in Eq.~\eqref{Eq:LeffbsgammaNC} and the loop functions $F_Z(r_p)$ and $F_Z'(r_p)$ in Eqs.~\eqref{Eq:FunctionFZ},~\eqref{Eq:FunctionFZd} do not agree with the corresponding terms of equations $(23),~(24)$ and the loop function $F_1^\mathrm{NC}(r_\alpha)$ of Ref.~\cite{Handoko:1994xw}.},
\begin{align}
  F_\mathrm{ZZ}(r_p,w_p)
  &\equiv
  F_1(r_p) + F_2(r_p) + F_3(w_p)~,\\[3pt]
  F_\mathrm{Z}(r_p)
  &\equiv
  2F_1(r_p)~,\label{Eq:FunctionFZ}\\[3pt]
  F_\mathrm{Z}'(r_p)
  &\equiv
  \frac{1-r_p^2+2r_p\ln r_p}{(1-r_p)^3}~\label{Eq:FunctionFZd},
\end{align}
where $r_p \equiv (m_d^p/M_Z)^2$ and $w_p \equiv (m_d^p/M_h)^2$. The functions $F_1$, $F_2$, and $F_3$,
\begin{align}
  F_1(r_p)
  &\equiv
  \frac{4-9r_p+5r_p^3+6r_p(1-2r_p)\ln r_p}{12(1-r_p)^4}~,\\[3pt]
  F_2(r_p)
  &\equiv
  r_p\frac{-20+39r_p-24r_p^2+5r_p^3+6(-2+r_p)\ln r_p}{24(1-r_p)^4}~,\\[3pt]
  F_3(w_p)
  &\equiv
  -w_p\frac{-16+45w_p-36w_p^2+7w_p^3+6(-2+3w_p)\ln w_p}{24(1-w_p)^4}~,
\end{align}
come from the diagram Fig.~\ref{Fig:bsgamma}$(b)$ where the exchanged particles are $Z$, $\chi_0$ and $h$ respectively.
The functions $f_\mathrm{ZZ}$, $f_\mathrm{Z}$ in Eq.~\eqref{Eq:LeffbsgammaastNC} are obtained as follows:
\begin{align}
  f_\mathrm{ZZ}(r_p,w_p)
  &\equiv
  f_1(r_p) + f_2(r_p) + f_3(w_p)~,\\[3pt]
  f_\mathrm{Z}(r_p)
  &\equiv
  2f_1(r_p)~,
\end{align}
where
\begin{align}
  f_1(r_p)
  &\equiv
  \frac{2+27r_p-54r_p^2+25r_p^3-6(2-9r_p+6r_p^2)\ln r_p}{18(1-r_p)^4}~,\\[3pt]
  f_2(r_p)
  &\equiv
  r_p\frac{-16+45r_p-36r_p^2+7r_p^3+6(-2+3r_p)\ln r_p}{36(1-r_p)^4}~,\\[3pt]
  f_3(w_p)
  &\equiv
  f_2(w_p)~.
\end{align}
The effective Lagrangians for the $b\rightarrow sg^{(*)}$ process can be obtained by replacing the external photon line which attached to quarks with gluon line in Fig.~\ref{Fig:bsgamma}.
They are obtained as follows:
\begin{align}
  \mathcal{L}_\mathrm{Eff}^\mathrm{CC}(b\rightarrow sg)
  &=
  -\frac{G_Fg_s}{8\sqrt{2}\pi^2}
  \sum_{i=c,t}\lambda_{sb}^i
  F_u(x_i)
  \overline{s}\sigma_{\mu\nu}\left(m_bR+m_sL\right)\frac{\lambda^a}{2}bG^{a\mu\nu}~,
  \label{Eq:LeffbsgCC}\\[3pt]
  \mathcal{L}_\mathrm{Eff}^\mathrm{uv}(b\rightarrow sg)
  &=
  \frac{G_Fg_s}{8\sqrt{2}\pi^2}\cdot\frac{2}{3}
  Z_\mathrm{NC}^{sb}
  \overline{s}\sigma_{\mu\nu}\left(m_bR+m_sL\right)\frac{\lambda^a}{2}bG^{a\mu\nu}~,
  \label{Eq:Leffbsguv}\\[3pt]
  \mathcal{L}_\mathrm{Eff}^\mathrm{NC}(b\rightarrow sg)
  &=
  \frac{G_Fg_s}{8\sqrt{2}\pi^2}
  \sum_{p=d,s,b}
  \Bigl\{
  Z_\mathrm{NC}^{sp}Z_\mathrm{NC}^{pb}F_\mathrm{ZZ}(r_p,w_p)\Bigr.
  \nonumber\\[3pt]
  &\phantom{=+}\Bigl.
  +Z_\mathrm{NC}^{sb}Q_ds_w^2(\delta^{sp}+\delta^{pb})F_\mathrm{Z}(r_p)
  \Bigr\}
  \overline{s}\sigma_{\mu\nu}\left(m_bR+m_sL\right)\frac{\lambda^a}{2}bG^{a\mu\nu}\nonumber\\[3pt]
  &\phantom{=}
  -\frac{G_Fg_s}{4\sqrt{2}\pi^2}
  \sum_{p=s,b}
  Z_\mathrm{NC}^{sb}Q_ds_w^2F_\mathrm{Z}'(r_p)
  \overline{s}\sigma_{\mu\nu}\left(\delta^{pb}m_bR+\delta^{sp}m_sL\right)\frac{\lambda^a}{2}bG^{a\mu\nu}~,\label{Eq:LeffbsgNC}
\end{align}
and
\begin{align}
  \mathcal{L}_\mathrm{Eff}^\mathrm{CC}(b\rightarrow sg^*)
  &=
  -\frac{G_Fg_s}{8\sqrt{2}\pi^2}
  \sum_{i=c,t}\lambda_{sb}^i
  f_u(x_i)
  \overline{s}\gamma_\nu L\frac{\lambda^a}{2}b \partial_\mu G^{a\mu\nu}~,\label{Eq:LeffbsgastCC}\\[3pt]
  \mathcal{L}_\mathrm{Eff}^\mathrm{uv}(b\rightarrow sg^*)
  &=
  -\frac{G_Fg_s}{8\sqrt{2}\pi^2}
  Z_\mathrm{NC}^{sb}
  \left(-\frac{2}{9}+\frac{4}{3}\ln x_u\right)
  \overline{s}\gamma_\nu L\frac{\lambda^a}{2}b \partial_\mu G^{a\mu\nu}~,\\[3pt]
  \mathcal{L}_\mathrm{Eff}^\mathrm{NC}(b\rightarrow sg^*)
  &=
  \frac{G_Fg_s}{8\sqrt{2}\pi^2}
  \sum_{p=d,s,b}
  \Bigl\{
  Z_\mathrm{NC}^{sp}Z_\mathrm{NC}^{pb}f_\mathrm{ZZ}(r_p,w_p)
  \nonumber\\[3pt]
  &\phantom{=+}
  +Z_\mathrm{NC}^{sb}Q_ds_w^2\left(\delta^{sp}+\delta^{pb}\right)f_\mathrm{Z}(r_p)
  \Bigr\}\overline{s}\gamma_\nu L\frac{\lambda^a}{2}b \partial_\mu G^{a\mu\nu}~.\label{Eq:LeffbsgastNC}
\end{align}

\section{Analysis of $B_s$-$\overline{B_s}$ mass difference, $B_s\rightarrow\mu^+\mu^-$, $\bar{B}\rightarrow X_s\gamma$ processes and violation of the CKM unitarity}
\label{sec:4}
In this section, we will make numerical calculations for the mass difference of the $B_s$ meson $\Delta M_{B_s}$, the branching ratio of the $B_s\rightarrow\mu^+\mu^-$ and the branching ratio of the  inclusive radiative decay of the $\bar{B}$ meson $\bar{B}\rightarrow X_s\gamma$ in the model with VLQ.
In addition to these processes, we consider the constraint from Eq.~\eqref{Eq:UnitarityRelation}.
We will use the new physics parameters defined as
\begin{align}
  r_{sb} \equiv
  \left|
  \frac{Z_\mathrm{NC}^{sb}}{\lambda_{sb}^t}
  \right|~,\quad
  \theta_{sb} \equiv
  \arg\left[
  \frac{Z_\mathrm{NC}^{sb}}{\lambda_{sb}^t}
  \right]~,\label{Eq.rtheta}
\end{align}
in the following computations.
\subsection{$B_s$-$\overline{B_s}$ mass difference}
We can obtain the mass difference of the $B_s$ meson in the model with VLQ as \cite{Branco:1992wr}-\cite{Barenboim:1997pf}:
\begin{align}
  \Delta M_{B_s}
  =
  \frac{G_F}{\sqrt{2}}\frac{\alpha_{em}}{6\pi s_w^2}
  \eta_B B_sf_{B_s}^2m_{B_s}
  \left|S_0(x_t)\right|
  \left|\lambda_{sb}^t\right|^2
  \left|\Delta_1(r_{sb},\theta_{sb})\right|~,
  \label{Eq:MassDifference}
\end{align}
where $\eta_B$, $B_s$ and $f_{B_s}$ represent the QCD factor, the bag parameter of $B_s$ meson and the $B_s$ meson decay constant respectively.
Here we use the QCD correction of the SM.
The numerical values for the parameters in Eq.~\eqref{Eq:MassDifference} are shown in Table~\ref{tab:ImputParameters}.
The function $\Delta_1(r_{sb},\theta_{sb})$ is given below,
\begin{align}
  \left|\Delta_1(r_{sb},\theta_{sb})\right|
  &=
  \left[
  1 + \frac{16Y_0(x_t)}{S_0(x_t)}r_{sb}\cos\theta_{sb}
  +\left\{
  \left|
  \frac{8Y_0(x_t)}{S_0(x_t)}
  \right|^2
  -\frac{8\pi s_w^2}{\alpha_{em}S_0(x_t)}\cos2\theta_{sb}
  \right\}r_{sb}^2\right.\nonumber\\[3pt]
  &\phantom{=[}\left.
  -\frac{64\pi s_w^2}{\alpha_{em}S_0(x_t)}\frac{Y_0(x_t)}{S_0(x_t)}r_{sb}^3\cos\theta_{sb}
  +\left|
  \frac{4\pi s_w^2}{\alpha_{em}S_0(x_t)}
  \right|^2r_{sb}^4
  \right]^{\frac{1}{2}}~.\label{Eq:Delta1}
\end{align}
We cannot use the SM value for the product of the CKM matrix elements $|\lambda_{sb}^t|$ in the model with VLQ since the new physics parameters $r_{sb}$ and $\theta_{sb}$ affect the determination of the CKM matrix elements. Instead we determine the $|\lambda_{sb}^t|$ by using Eq.~\eqref{Eq:MassDifference} in the following computations.
Therefore the $|\lambda_{sb}^t|$ is obtained as the function with respect to the new physics parameters $r_{sb}$ and $\theta_{sb}$.
\subsection{Branching ratio of $B_s\rightarrow\mu^+\mu^-$}
The branching ratio of the $B_s\rightarrow\mu^+\mu^-$ process in the model with VLQ is given as follows:
\begin{align}
  &\phantom{=}\mathrm{Br}\left[B_s\rightarrow\mu^+\mu^-\right]_\mathrm{VLQ}\nonumber\\[3pt]
  &=\tau_{B_s}\frac{G_F^2}{16\pi}
  \left(\frac{\alpha_{em}}{\pi s_w^2}\right)^2
  \left|\eta_YY_0(x_t)\right|^2\left|f_{B_s}\right|^2
  m_{B_s}m_\mu^2
  \sqrt{1-\frac{4m_\mu^2}{m_{B_s}^2}}\left|\lambda_{sb}^t\right|^2
  \left|\Delta_2(r_{sb},\theta_{sb})\right|^2~,
  \label{Eq:BrBsmumu}
\end{align}
where $\eta_Y$ is the NLO QCD correction \cite{Buchalla:1995vs,Buras:2012ru}. The life time of the $B_s$ meson is denoted by $\tau_{B_s}$. These values are shown in Table~\ref{tab:ImputParameters}.
The function $\Delta_2(r_{sb},\theta_{sb})$ is given below,
\begin{align}
  \left|\Delta_2(r_{sb},\theta_{sb})\right|
  =
  \left[
  1-\frac{2\pi s_w^2}{\alpha_{em}Y_0(x_t)}r_{sb}\cos\theta_{sb}
  +\left\{\frac{\pi s_w^2}{\alpha_{em}Y_0(x_t)}\right\}^2r_{sb}^2
  \right]^\frac{1}{2}~.\label{Eq:Delta2}
\end{align}
\subsection{Branching ratio of $\bar{B}\rightarrow X_s\gamma$}
The $\bar{B}$ meson inclusive radiative decay $\bar{B}\rightarrow X_s\gamma$ is governed by the effective Hamiltonian at the $b$-quark mass scale $\mu=\mathcal{O}(m_b)$ \cite{Buchalla:1995vs,Grinstein:1990tj},
\begin{align}
  \mathcal{H}_\mathrm{Eff}(b\rightarrow s\gamma)
  =
  -\frac{G_F}{\sqrt{2}}\lambda_{sb}^t
  \left[
  \sum_{i=1}^6C_i(\mu)O_i(\mu)
  +C_{7\gamma}(\mu)O_{7\gamma}(\mu)
  +C_{8G}(\mu)O_{8G}(\mu)
  \right]~,\label{Eq:HEffbsgamma}
\end{align}
where $O_i$ and $C_i~(i=1\sim 6)$ denote the 4-Fermi operators and their Wilson coefficients respectively. The effective operators $O_{7\gamma}$ and $O_{8G}$ are given by,
\begin{align}
  O_{7\gamma}
  &=
  \frac{e}{8\pi^2}m_b\overline{s}\sigma_{\mu\nu}(1+\gamma_5)bF_A^{\mu\nu}~,\\[3pt]
  O_{8G}
  &=
  \frac{g_s}{8\pi^2}m_b\overline{s}\frac{\lambda^a}{2}\sigma_{\mu\nu}(1+\gamma_5)bG^{a\mu\nu}~.
\end{align}
In the calculation of the branching ratio for the $\bar{B}\rightarrow X_s\gamma$, it is convenient to introduce the so-called ``effective coefficients'' $C_i^{(0)eff}$ \cite{Ciuchini:1993ks,Buras:1993xp}.
The effective coefficient for the effective operator $O_{7\gamma}$ at the scale $\mu=\mathcal{O}(m_b)$ is given as \cite{Buchalla:1995vs,Buras:1993xp,Buras:1998raa}:
\begin{align}
  C_{7\gamma}^{(0)eff}(\mu)
  =
  \eta^\frac{16}{23}C_{7\gamma}^{(0)}(M_W)
  + \frac{8}{3}\left(\eta^\frac{14}{23}-\eta^\frac{16}{23}\right)C_{8G}^{(0)}(M_W)
  + C_2^{(0)}(M_W)\sum_{i=1}^{8}h_i\eta^{a_i},~\label{Eq:C7eff}
\end{align}
where $\eta = \alpha_s(M_W)/\alpha_s(\mu)$ with $\alpha_s = g_s^2/(4\pi)$ and
\begin{align}
  h_i
  &=\left(
  2.2996,~-1.0880,~-\frac{3}{7},~-\frac{1}{14},~-0.6494,~-0.0380,~-0.0185,~-0.0057
  \right)~,\\[3pt]
  a_i
  &=
  \left(
  \frac{14}{23},~\frac{16}{23},~\frac{6}{23},~-\frac{12}{23},~0.4086,~-0.4230,~-0.8994,~0.1456
  \right)~.
\end{align}
In Eq.~\eqref{Eq:C7eff}, the indices ``(0)'' mean the leading order contributions.
Since we do not take into account the running effect from the VLQ mass scale to the EW scale, we obtain the Wilson coefficients $C_{7\gamma}^{(0)}$ and $C_{8G}^{(0)}$ at the EW scale (taken as $M_W$) as:
\begin{align}
  C_{7\gamma}^{(0)}(M_W)
  &=
  C_{7\gamma}^\mathrm{SM}(M_W)
  +
  C_{7\gamma}^\mathrm{NP1}(M_W)
  +
  C_{7\gamma}^\mathrm{NP2}(M_W)~,\label{Eq:C7MW}\\[3pt]
  C_{8G}^{(0)}(M_W)
  &=
  C_{8G}^\mathrm{SM}(M_W)
  +
  C_{8G}^\mathrm{NP1}(M_W)
  +
  C_{8G}^\mathrm{NP2}(M_W)~,\label{Eq:C8MW}
\end{align}
where the Wilson coefficients,
\begin{align}
  C_{7\gamma}^\mathrm{SM}(M_W)
  &=
  -\frac{1}{2}\left[Q_uF_u(x_t)+F_W(x_t)\right]~,\\[3pt]
  C_{8G}^\mathrm{SM}(M_W)
  &=
  -\frac{1}{2}F_u(x_t)~,
\end{align}
come from the SM contributions in Eq.~\eqref{Eq:LeffbsgammaCC}. The Wilson coefficients
\begin{align}
  C_{7\gamma}^\mathrm{NP1}(M_W)
  &=
  C_{7\gamma}^\mathrm{NP1}(M_4)
  =
  \frac{Q_d}{24}\cdot\frac{Z_\mathrm{NC}^{sb}}{\lambda_{sb}^t}~,\\[3pt]
  C_{8G}^\mathrm{NP1}(M_W)
  &=
  C_{8G}^\mathrm{NP1}(M_4)
  =
  \frac{1}{24}\cdot\frac{Z_\mathrm{NC}^{sb}}{\lambda_{sb}^t}~,
\end{align}
are obtained from the VLQ contributions in Eqs.~\eqref{Eq:DipoleEW} and \eqref{Eq:DipoleGluon}.
The Wilson coefficients $C_{7\gamma}^\mathrm{NP2}$ and $C_{8G}^\mathrm{NP2}$
are given as follows:
\begin{align}
  C_{7\gamma}^\mathrm{NP2}(M_W)
  &=
  C_{7\gamma}^\mathrm{uv}(M_W)
  +
  C_{7\gamma}^\mathrm{NC}(M_W)~,\\[3pt]
  C_{8G}^\mathrm{NP2}(M_W)
  &=
  C_{8G}^\mathrm{uv}(M_W)
  +
  C_{8G}^\mathrm{NC}(M_W)~,
\end{align}
where
\begin{align}
  \begin{array}{ll}
  C_{7\gamma}^\mathrm{uv}(M_W)
  =
  \dfrac{1}{2}
  \left(
  \dfrac{2}{3}Q_u + \dfrac{5}{6}
  \right)\cdot\dfrac{Z_\mathrm{NC}^{sb}}{\lambda_{sb}^t}
  ~, &
  C_{7\gamma}^\mathrm{NC}(M_W)
  =
  \dfrac{Q_d}{3}\left(1-Q_ds_w^2\right)\cdot\dfrac{Z_\mathrm{NC}^{sb}}{\lambda_{sb}^t}~,\\
   & \\
  C_{8G}^\mathrm{uv}(M_W)
  =
  \dfrac{1}{3}\cdot\dfrac{Z_\mathrm{NC}^{sb}}{\lambda_{sb}^t}
  ~, &
  C_{8G}^\mathrm{NC}(M_W)
  =
  \dfrac{1}{3}\left(1-Q_ds_w^2\right)\cdot\dfrac{Z_\mathrm{NC}^{sb}}{\lambda_{sb}^t}~.
  \end{array}
\end{align}
The Wilson coefficients with the suffix ``uv'' come from the effective Lagrangian in Eq.~\eqref{Eq:Leffbsgammauv} whose origin is the violation of the CKM unitarity. The Wilson coefficients with the suffix ``NC'' are obtained from the effective Lagrangian in Eq.~\eqref{Eq:LeffbsgammaNC} by taking the limit $r_p\rightarrow 0$ and $w_p\rightarrow 0$.
Here we neglect $\mathcal{O}(Z_\mathrm{NC}^2)$ terms.
The Wilson coefficients $C_{8G}^\mathrm{uv}$ and $C_{8G}^\mathrm{NC}$ can be obtained from the effective Lagrangian corresponding to the $b\rightarrow sg$ diagrams.

In our numerical calculation, we use the NLO expression for the branching ratio $\mathrm{Br}[\bar{B}\rightarrow X_s\gamma]$ given as \cite{Chetyrkin:1996vx}:
\begin{align}
  \mathrm{Br}[\bar{B}\rightarrow X_s\gamma]
  =
  \mathrm{Br}[\bar{B}\rightarrow X_ce\overline{\nu_e}]_\mathrm{Exp}
  \cdot
  R_\mathrm{quark}(\delta)
  \left(1-\frac{\delta_{sl}^{NP}}{m_b^2}+\frac{\delta_{rad}^{NP}}{m_b^2}\right)~,\label{Eq:BrBXsgamma}
\end{align}
where $\delta_{sl}^{NP}$ and $\delta_{rad}^{NP}$ are non-perturbative correction for the semi-leptonic and radiative $\bar{B}$ meson decay rates, respectively. The quantity $R_\mathrm{quark}$ at NLO is summarized in Ref.~\cite{Chetyrkin:1996vx} as:
\begin{align}
  R_\mathrm{quark}(\delta)
  =
  \frac{\Gamma[b\rightarrow X_s\gamma]^{E_\gamma>(1-\delta)E_\gamma^{max}}}{\Gamma[b\rightarrow X_ce\overline{\nu_e}]}
  =
  \frac{\left|\lambda_{sb}^t\right|^2}{\left|V_\mathrm{CKM}^{cb}\right|^2}
  \frac{6\alpha_{em}}{\pi g(z)}F(z)
  \left\{|D|^2 + A(\delta)\right\}~.\label{Eq:RquarkNLO}
\end{align}
The function $g(z)$ with $z=m_{c,pole}^2/m_{b,pole}^2$ corresponds to the phase space factor for the semi-leptonic decay. The function $F(z)$ contains the NLO correction for the semi-leptonic decay and the difference between the pole mass and $\mathrm{\overline{MS}}$ mass of the $b$-quark.
The $\delta$ is the lower cut on the photon energy in the bremsstrahlung correction:
\begin{align}
  E_\gamma>(1-\delta)E_\gamma^{max}
  \equiv
  (1-\delta)\frac{m_b}{2}~.
\end{align}
The term $A(\delta)$ originates from the bremsstrahlung corrections and the virtual corrections \cite{Chetyrkin:1996vx}-\cite{Pott:1995if},
\begin{align}
  A
  =
  \left\{
  e^{-\frac{\alpha_s(\mu_b)}{3\pi}(7+2\ln\delta)\ln\delta}-1
  \right\}\left|C_{7\gamma}^{(0)eff}(\mu_b)\right|^2
  +\frac{\alpha_s(\mu_b)}{\pi}
  \sum_{\substack{i,j=1 \\ i\leq j}}^8
  C_i^{(0)eff}(\mu_b)C_j^{(0)eff}(\mu_b)f_{ij}(\delta)~,
\end{align}
where the functions $f_{ij}(\delta)$ can be found in Ref.~\cite{Chetyrkin:1996vx}.
The term $|D|^2$ in Eq.~\eqref{Eq:RquarkNLO} is constituted by the NLO Wilson coefficient for $O_{7\gamma}$ and the virtual corrections for $b\rightarrow s\gamma$ \cite{Chetyrkin:1996vx}-\cite{Greub:1996tg}. Here $D$ is defined as,
\begin{align}
  D
  =
  C_{7\gamma}^{(0)eff}(\mu_b)
  +
  \frac{\alpha_s(\mu_b)}{4\pi}
  \left[
  C_{7\gamma}^{(1)eff}(\mu_b)
  +
  \sum_{i=1}^8C_i^{(0)eff}(\mu_b)
  \left\{r_i + \gamma_{i7}^{(0)eff}\ln\frac{m_b}{\mu_b}
  \right\}
  \right]~,\label{Eq:D}
\end{align}
where $C_{7\gamma}^{(1)eff}(\mu_b)$, $r_i$ and $\gamma_{i7}^{(0)eff}$ can be found in Ref.~\cite{Chetyrkin:1996vx}.

In our numerical calculation, we take $\mu_b = m_b$, $E_\gamma >1.6$ GeV and neglect the $\mathcal{O}(\alpha_s)$ correction to the new physics contributions. Therefore the Wilson coefficients $C_{7\gamma,8G}^\mathrm{NP1}$ and $C_{7\gamma,8G}^\mathrm{NP2}$ are only included in the first term in Eq.~\eqref{Eq:D}.
\subsection{Violation of CKM Unitarity}
\begin{figure}[tbp]
  \centering
  \includegraphics[width=8.0cm]{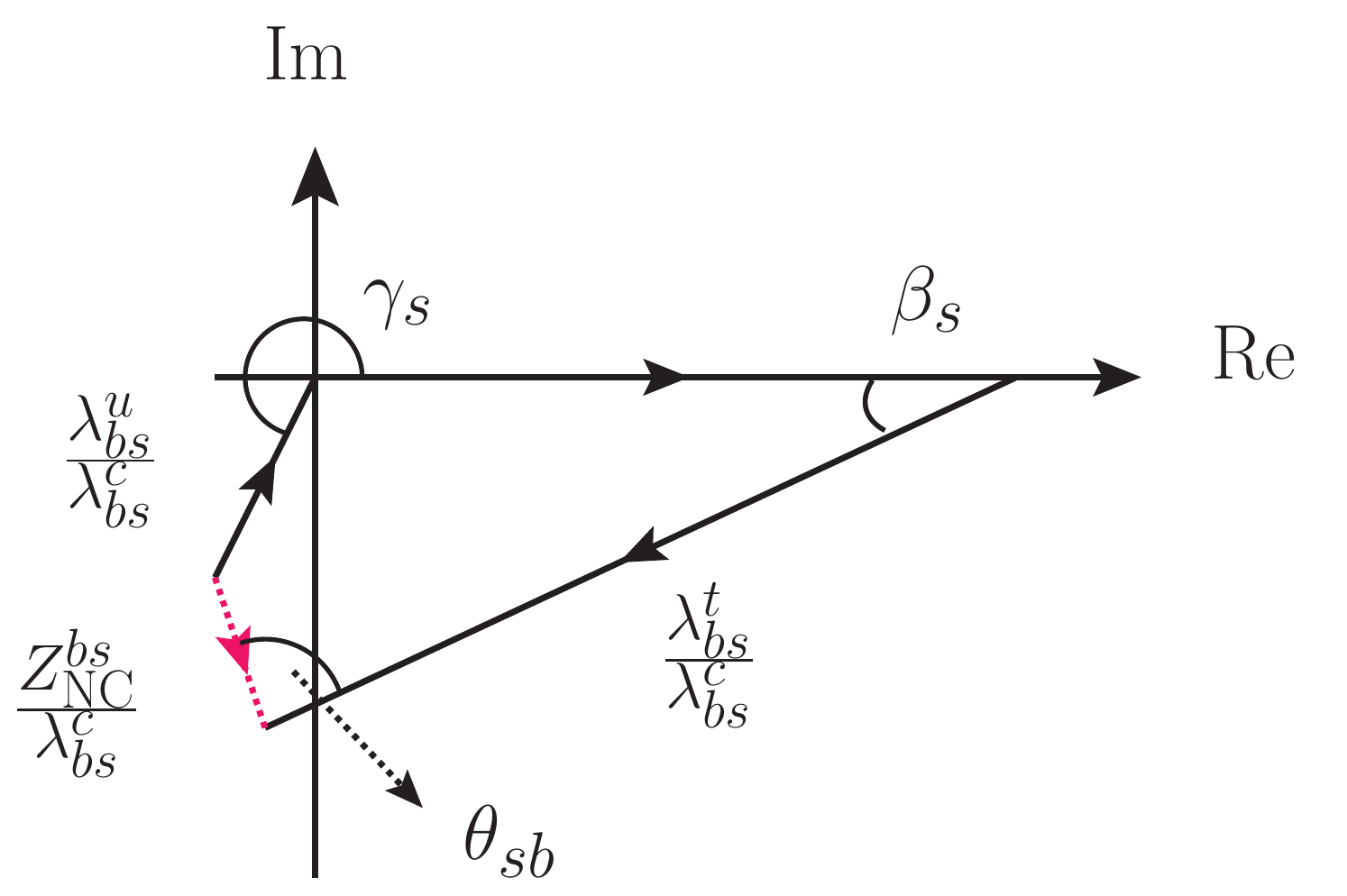}
  \caption{The violation of CKM unitarity Eq.~\eqref{Eq:UnitarityRelationbs} in the complex plane.
  We multiply the relation in Eq.~\eqref{Eq:UnitarityRelationbs} by a factor of $1/\lambda_{bs}^c$.}
  \label{Fig:Quadrangle}
\end{figure}

The violation of CKM unitarity is shown in Eq.~\eqref{Eq:UnitarityRelation}. For $p=b,q=s$, we obtain the following relation:
\begin{align}
  \lambda_{bs}^u + \lambda_{bs}^c + \lambda_{bs}^t
  =
  Z_\mathrm{NC}^{bs}~.\label{Eq:UnitarityRelationbs}
\end{align}
This relation can be rewritten as follows:
\begin{align}
  \left|\frac{\lambda_{bs}^c}{\lambda_{bs}^t}\right|^2
  \left(
  1-2\left|\frac{\lambda_{bs}^u}{\lambda_{bs}^c}\right|\cos\gamma_{s}
  +\left|\frac{\lambda_{bs}^u}{\lambda_{bs}^c}\right|^2
  \right)
  =
  1-2r_{sb}\cos\theta_{sb}+r_{sb}^2~,\label{Eq:UnitarityRelationbs2}
\end{align}
where we define
\begin{align}
  \gamma_{s} \equiv
  \arg\left[
  -\frac{\lambda_{bs}^u}{\lambda_{bs}^c}
  \right]~.
\end{align}
The relation in Eq.~\eqref{Eq:UnitarityRelationbs} leads to a quadrangle in the complex plane as shown in Fig.~\ref{Fig:Quadrangle}.

\subsection{Numerical Analyses}

In the following numerical analyses, we obtain the constraints on FCNC couplings by using the current experimental data of rare $B$ decays $B_s\rightarrow\mu^+\mu^-$ and $\bar{B}\rightarrow X_s\gamma$.
We also take account of the quadrangle constraint Eq.~\eqref{Eq:UnitarityRelationbs2} and Fig.~\ref{Fig:Quadrangle}.
The values of the input parameters used in the numerical analyses are shown in Table~\ref{tab:ImputParameters}.
{\renewcommand\arraystretch{1.8}
\begin{table}[h]
  \begin{center}
    \caption{The values of input parameters}
    \footnotesize
    \begin{tabular}{cc|cc}
      \multicolumn{4}{c}{}\\
      \hline\hline
      $\alpha_{em}^{-1}(m_b\sim M_W)$ & $130.3\pm 2.3$ \cite{Chetyrkin:1996vx} &
      $\alpha_s(M_Z)$ & $0.1181\pm 0.0011$ \cite{Patrignani:2016xqp} \\
      $M_W$ & $80.385\pm 0.015$ GeV \cite{Patrignani:2016xqp} &
      $M_Z$ & $91.1876\pm 0.0021$ GeV \cite{Patrignani:2016xqp}\\
      $G_F$ & $1.16638\times 10^{-5}$ $\mathrm{GeV}^{-2}$ \cite{Patrignani:2016xqp}&
      $\sin\theta_w$ & $0.23129$ \cite{Patrignani:2016xqp}\\
      $m_{c,\overline{\mathrm{MS}}}$ & $1.28\pm 0.03$ GeV \cite{Patrignani:2016xqp} &
      $m_{b,\overline{\mathrm{MS}}}$ & $4.18^{+0.04}_{-0.03}$ GeV \cite{Patrignani:2016xqp}\\
      $m_{t,pole}$ & $173.5\pm 1.1$ GeV \cite{Patrignani:2016xqp} &
      $m_\mu$ & $105.6584$ MeV \cite{Patrignani:2016xqp}\\
      $m_B$ & $5.27963\pm 0.00015$ GeV \cite{Patrignani:2016xqp} &
      $m_{B_s}$ & $5.36689\pm 0.00019$ GeV \cite{Patrignani:2016xqp}\\
      $\tau_{B_s}$ & $(1.505\pm 0.005)\times 10^{-12}$ s \cite{Patrignani:2016xqp} &
      $\Delta M_{B_s}$ & $(1.1688\pm 0.0014)\times 10^{-8}$ MeV \cite{Patrignani:2016xqp}\\
      $\mathrm{Br}[\bar{B}\rightarrow X_c e\overline{\nu_e}]_\mathrm{Exp}$ & $(10.1\pm 0.4)\times 10^{-2}$ \cite{Patrignani:2016xqp} &
      $\eta_Y$ & $1.0113$ \cite{Buras:2012ru} \\
      $\eta_B$ & $0.5510\pm 0.0022$ \cite{Buchalla:1995vs} &
      $f_{B_s}$ & $225.1\pm 1.5\pm 2.0$ MeV \cite{Charles:2004jd}\\
      $B_s$ & $1.320\pm 0.016\pm 0.030$ \cite{Charles:2004jd} &
      $V_{us}$ & $0.22508^{+0.00030}_{-0.00028}$ \cite{Charles:2004jd} \\
      $V_{ub}$ & $0.003715^{+0.000060}_{-0.000060}$ \cite{Charles:2004jd} &
      $V_{cs}$ & $0.973471^{+0.000067}_{-0.000067}$ \cite{Charles:2004jd}\\
      $V_{cb}$ & $0.04181^{+0.00028}_{-0.00060}$ \cite{Charles:2004jd} &
       &
    \end{tabular}\label{tab:ImputParameters}
  \end{center}
\end{table}
}

\begin{figure}[tbp]
  \centering
  \includegraphics[width=8.0cm]{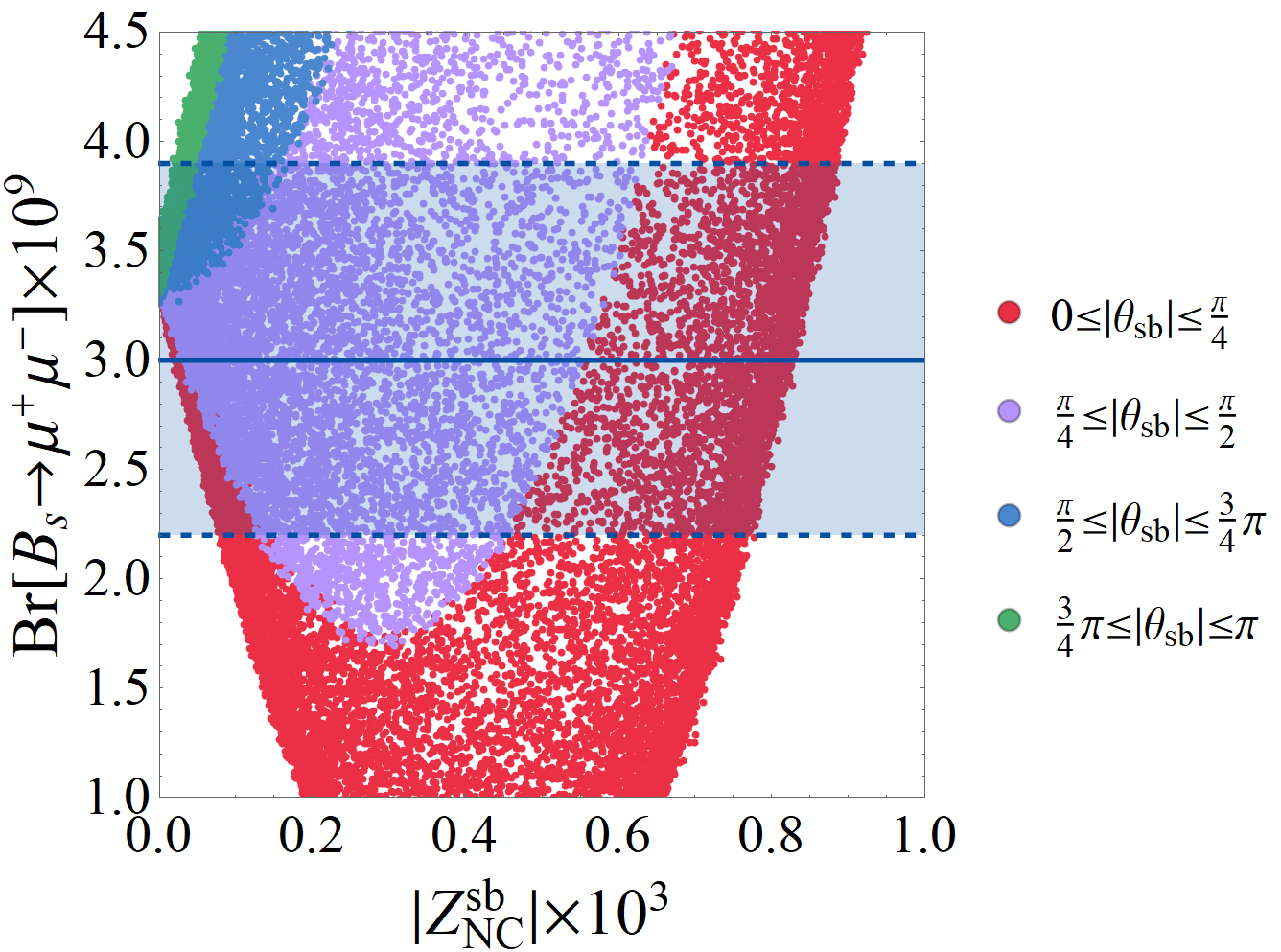}
  \caption{The dependence of $\mathrm{Br}[B_s\rightarrow\mu^+\mu^-]_\mathrm{VLQ}$ on the FCNC coupling $\left|Z_\mathrm{NC}^{sb}\right|$.
The different colors of the dots represent the different ranges for the value of $|\theta_{sb}|$.
All the colored region satisfy the quadrangle constraint Eq.~\eqref{Eq:UnitarityRelationbs2} with $0\leq\gamma_s\leq2\pi$.
The experimentally allowed range of the branching ratio in Eq.~\eqref{Eq:ExpBsmumu} is shown as the blue shaded region.
The horizontal solid line corresponds to the central value of the experimental branching ratio in Eq.~\eqref{Eq:ExpBsmumu}.
}
  \label{Fig:ConstraintBsmumu}
\end{figure}

At first we analyze the branching ratio of $B_s\rightarrow\mu^+\mu^-$ process by using the expression in Eq.~\eqref{Eq:BrBsmumu}.
Note that the branching ratio depends on the new physics parameters $r_{sb}$ and $\cos\theta_{sb}$.
We equate $\mathrm{Br}[B_s\rightarrow\mu^+\mu^-]_\mathrm{VLQ}$ in Eq.~\eqref{Eq:BrBsmumu} with the experimental value,
\begin{align}
  \mathrm{Br}[B_s\rightarrow\mu^+\mu^-]_\mathrm{VLQ}
  =
  \mathrm{Br}[B_s\rightarrow\mu^+\mu^-]_\mathrm{Exp}~.
\end{align}
As the experimental value, we adopt the branching ratio measured by LHCb \cite{Aaij:2017vad},
\begin{align}
  \mathrm{Br}[B_s\rightarrow\mu^+\mu^-]_\mathrm{Exp} &= \left(3.0\pm0.6^{+0.3}_{-0.2}\right)\times 10^{-9}~.\label{Eq:ExpBsmumu}
\end{align}
In Fig.~\ref{Fig:ConstraintBsmumu}, we show the dependence of  $\mathrm{Br}[B_s\rightarrow\mu^+\mu^-]_\mathrm{VLQ}$ on the absolute value of the FCNC coupling $\left|Z_\mathrm{NC}^{sb}\right|$.
The different colors of the dots in the scattered plots represent the different ranges for the value of the new physics parameter $|\theta_{sb}|$.
All the colored region satisfy the quadrangle constraint Eq.~\eqref{Eq:UnitarityRelationbs2} with $0\leq\gamma_s\leq2\pi$.
Note that the expression of the branching ratio and quadrangle constraint depend on $\theta_{sb}$ through its cosine and the plotted regions do not depend on the sign of $\theta_{sb}$.
The experimentally allowed range of the branching ratio in Eq.~\eqref{Eq:ExpBsmumu} is shown as the blue shaded region.
The horizontal solid line corresponds to the central value of the experimental branching ratio in Eq.~\eqref{Eq:ExpBsmumu}.
In Fig.~\ref{Fig:ConstraintBsmumu}, as $\left|Z_\mathrm{NC}^{sb}\right|$ approaches zero, $\mathrm{Br}[B_s\rightarrow\mu^+\mu^-]_\mathrm{VLQ}$ comes close to the SM prediction \cite{Buras:2012ru},
\begin{align}
  \mathrm{Br}[B_s\rightarrow\mu^+\mu^-]_\mathrm{SM} = \left(3.23\pm 0.27\right)\times 10^{-9}~.
\end{align}
As $\left|Z_\mathrm{NC}^{sb}\right|$ increases from zero to $3\times 10^{-4}$, $\mathrm{Br}[B_s\rightarrow\mu^+\mu^-]_\mathrm{VLQ}$ decreases for $|\theta_{sb}|<\pi/2$,
while it increases for $\pi/2<|\theta_{sb}|<\pi$.
As $\left|Z_\mathrm{NC}^{sb}\right|$ becomes larger,
$\mathrm{Br}[B_s\rightarrow\mu^+\mu^-]_\mathrm{VLQ}$ increases regardless of the range of the $|\theta_{sb}|$ since the third term in Eq.~\eqref{Eq:Delta2} is dominant.
The dependence on $|\theta_{sb}|$ for the smaller $\left|Z_\mathrm{NC}^{sb}\right|$ can be also understood from the Eq.~\eqref{Eq:Delta2} since the coefficient of the term which is linear to $r_{sb}$ is proportional to $\cos\theta_{sb}$.

\begin{figure}[tbp]
  \begin{center}
 \begin{minipage}{0.47\hsize}
  \begin{center}
   \includegraphics[width=7.8cm]{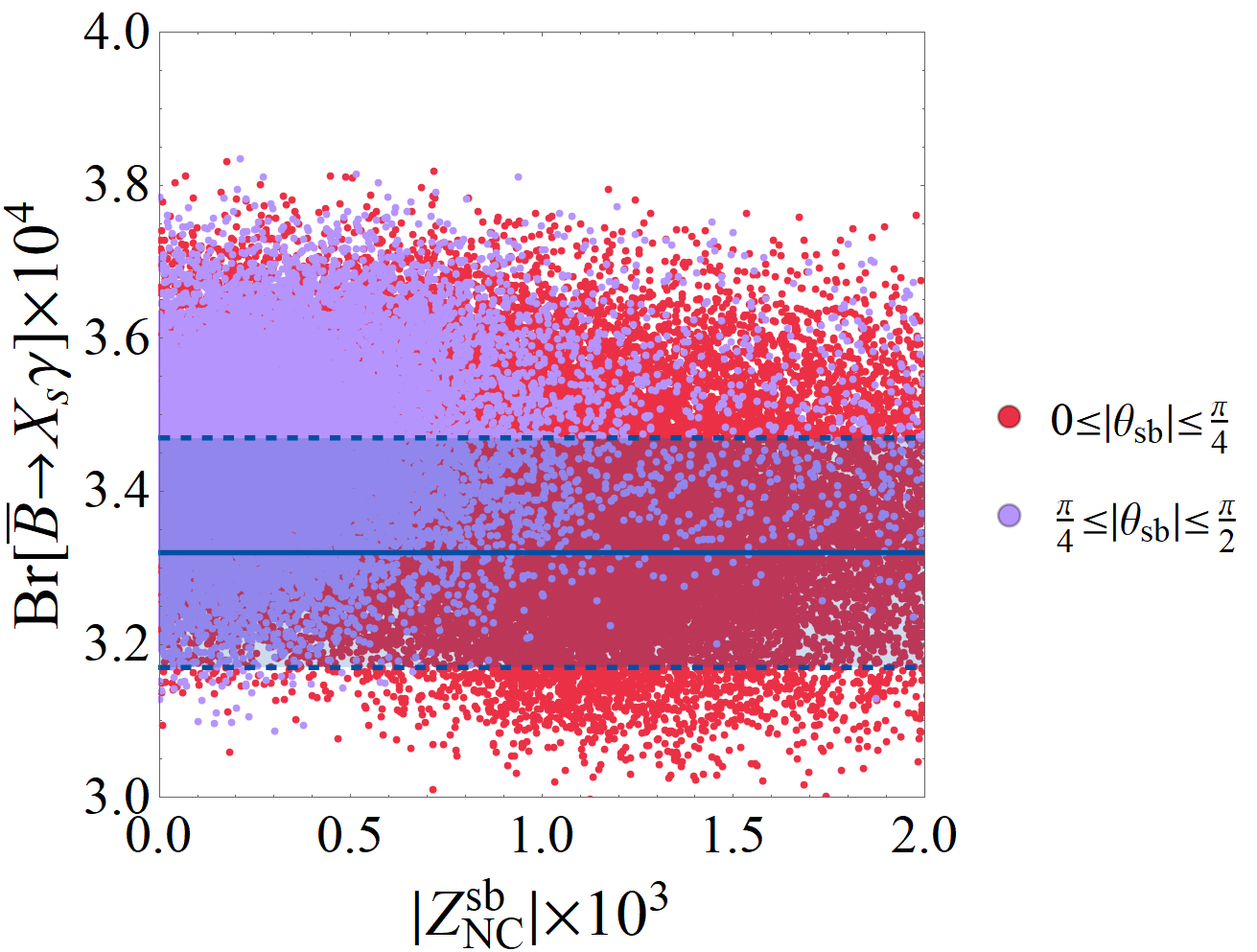}
  \end{center}
 \end{minipage}
 \hspace{0.5cm}
 \begin{minipage}{0.47\hsize}
  \begin{center}
   \includegraphics[width=7.8cm]{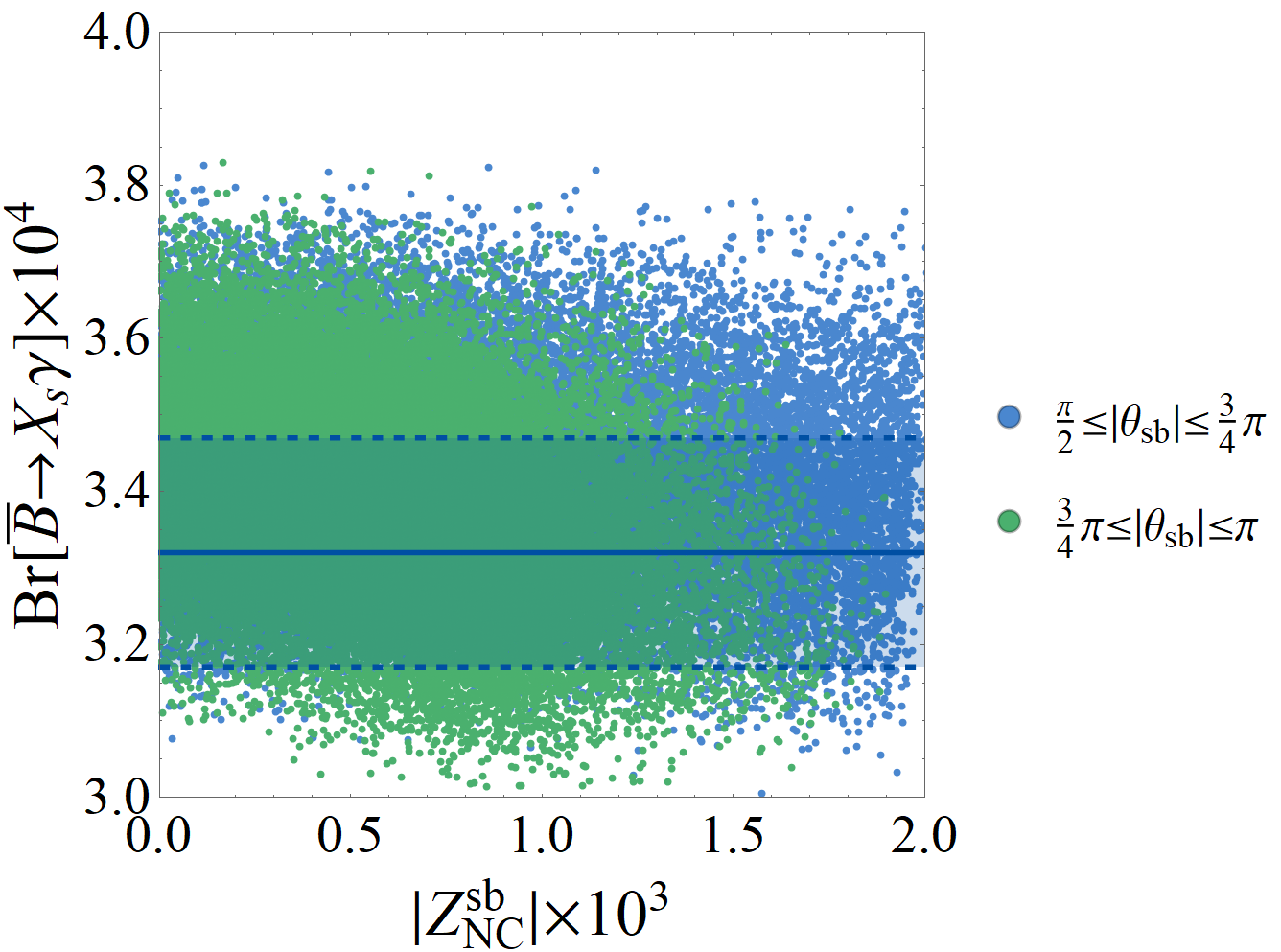}
  \end{center}
 \end{minipage}
 \caption{The dependence of $\mathrm{Br}[\bar{B}\rightarrow X_s\gamma]_\mathrm{VLQ}$ on $|Z_\mathrm{NC}^{sb}|$.
The different colors of the dots represent the different ranges for the value of $|\theta_{sb}|$.
All the colored region satisfy the quadrangle constraint Eq.~\eqref{Eq:UnitarityRelationbs2} with $0\leq\gamma_s\leq2\pi$.
The experimentally allowed range of the branching ratio in Eq.~\eqref{Eq:ExpBXsgamma} is shown as the blue shaded region.
The horizontal solid line corresponds to the central value of the experimental branching ratio in Eq.~\eqref{Eq:ExpBXsgamma}.}
 \label{Fig:ConstraintBXsgamma}
 \end{center}
\end{figure}

Next we analyze the branching ratio of $\bar{B}\rightarrow X_s\gamma$ process by using the expression in Eq.~\eqref{Eq:BrBXsgamma}.
Here we denote the branching ratio in the model with VLQ as $\mathrm{Br}[\bar{B}\rightarrow X_s\gamma]_\mathrm{VLQ}$.
The new physics parameters $r_{sb}$ and $\theta_{sb}$ are included in the Wilson coefficients in Eqs.~\eqref{Eq:C7MW} and \eqref{Eq:C8MW}. In order to obtain constraints on $r_{sb}$ and $\theta_{sb}$, we take account of the current average \cite{Amhis:2016xyh},
\begin{align}
  \mathrm{Br}[\bar{B}\rightarrow X_s\gamma]_\mathrm{Exp}
  =
  \left(3.32\pm0.15\right)\times 10^{-4}~,\label{Eq:ExpBXsgamma}
\end{align}
of experimental data \cite{Aubert:2007my}-\cite{Chen:2001fja}.
In Fig.~\ref{Fig:ConstraintBXsgamma}, we show the dependence of $\mathrm{Br}[\bar{B}\rightarrow X_s\gamma]_\mathrm{VLQ}$ on $|Z_\mathrm{NC}^{sb}|$.
The different colors of the dots in the scattered plots represent the different ranges for the value of the new physics parameter $|\theta_{sb}|$.
All the colored region satisfy the quadrangle constraint Eq.~\eqref{Eq:UnitarityRelationbs2} with $0\leq\gamma_s\leq2\pi$.
The experimentally allowed range of the branching ratio in Eq.~\eqref{Eq:ExpBXsgamma} is shown as the blue shaded region.
The horizontal solid line corresponds to the central value of the experimental branching ratio in Eq.~\eqref{Eq:ExpBXsgamma}.
In Fig.~\ref{Fig:ConstraintBXsgamma},
as $\left|Z_\mathrm{NC}\right|$ approaches zero, the value of $\mathrm{Br}[\bar{B}\rightarrow X_s\gamma]_\mathrm{VLQ}$ comes close to that of the SM prediction at NNLO accuracy \cite{Misiak:2015xwa},
\begin{align}
  \mathrm{Br}[\bar{B}\rightarrow X_s\gamma]_\mathrm{SM}
  =
  \left(3.36\pm 0.23\right)\times 10^{-4}~.
\end{align}
We note that the number of the purple colored dots is much less than that of the red colored ones, since the quadrangle constraint for $\pi/4\leq\theta_{sb}\leq\pi/2$ is tighter than that for $0\leq\theta_{sb}\leq\pi/4$.
For the smaller $\left|Z_\mathrm{NC}^{sb}\right|$, the filled regions with colored dots are the almost same as each other. Thus $\mathrm{Br}[B\rightarrow X_s\gamma]_\mathrm{VLQ}$ depends on $|\theta_{sb}|$ weakly compared with $\mathrm{Br}[B_s\rightarrow\mu^+\mu^-]_\mathrm{VLQ}$.

\begin{figure}[tbp]
  \begin{center}
 \begin{minipage}{0.51\hsize}
  \begin{center}
   \includegraphics[width=8.4cm]{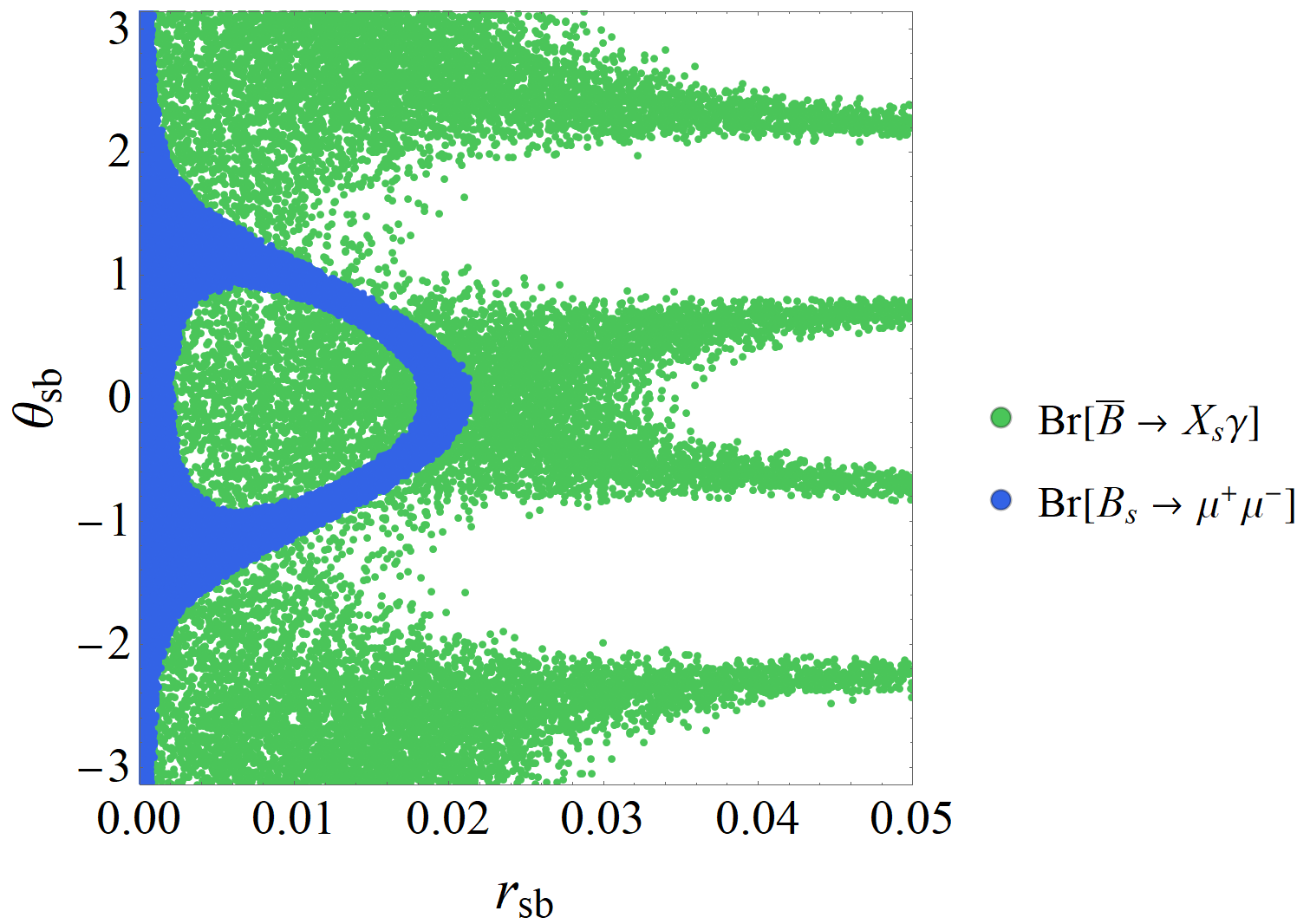}
  \end{center}
 \end{minipage}
 \hspace{0.5cm}
 \begin{minipage}{0.43\hsize}
  \begin{center}
   \includegraphics[width=6.8cm]{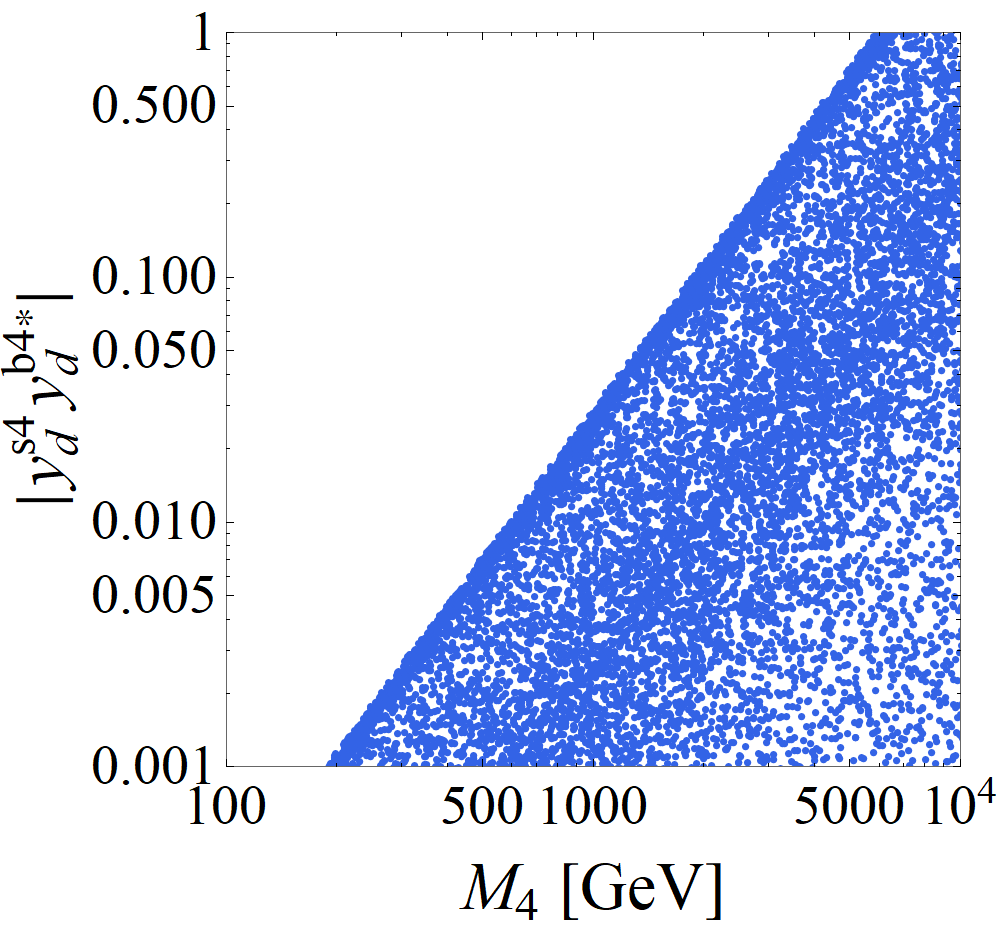}
  \end{center}
 \end{minipage}
 \caption{(Left) : The region allowed by the experimental data for the parameter $r_{sb}$ and $\theta_{sb}$.
The blue dots satisfy both the constraint from the $\mathrm{Br}[B_s\rightarrow\mu^+\mu^-]_\mathrm{Exp}$ and the quadrangle constraint Eq.~\eqref{Eq:UnitarityRelationbs2} with $0\leq\gamma_s\leq2\pi$.
The green dots satisfy both the constraint from the $\mathrm{Br}[\bar{B}\rightarrow X_s\gamma]_\mathrm{Exp}$ and the quadrangle constraint.
(Right) : The constraint on the VLQ mass $M_4$ and the product of the Yukawa coupling $|y_d^{s4}y_d^{b4*}|$. The blue region satisfies the constraints from the $\mathrm{Br}[B_s\rightarrow\mu^+\mu^-]_\mathrm{Exp}$ and the quadrangle constraint Eq.~\eqref{Eq:UnitarityRelationbs2}.}
 \label{Fig:ConstraintConv}
 \end{center}
\end{figure}

In the left figure of Fig.~\ref{Fig:ConstraintConv}, we show the region allowed by the experimental data for the parameter $r_{sb}$ and $\theta_{sb}$.
The blue dots satisfy both the constraint from the $\mathrm{Br}[B_s\rightarrow\mu^+\mu^-]_\mathrm{Exp}$ and the quadrangle constraint Eq.~\eqref{Eq:UnitarityRelationbs2} with $0\leq\gamma_s\leq2\pi$.
The green dots satisfy both the constraint from the $\mathrm{Br}[\bar{B}\rightarrow X_s\gamma]_\mathrm{Exp}$ and the quadrangle constraint.
The values of $r_{sb}$ and $\theta_{sb}$ in the region where the blue and  green region overlap each other satisfy all the three constraints.
The blue region has the shape of a ring.
The region inside the ring is excluded because $r_{sb}$ and $\theta_{sb}$ in this region leads to the predictions of $\mathrm{Br}[B_s\rightarrow\mu^+\mu^-]$ smaller than the experimental value.
One finds that the stringent constraint on the parameters $r_{sb}$ and $\theta_{sb}$ comes from the $\mathrm{Br}[B_s\rightarrow\mu^+\mu^-]_\mathrm{Exp}$.

Using the definition of $Z_\mathrm{NC}$ in Eq.~\eqref{Eq:ZNC}, we obtain the constraint on the VLQ mass $M_4$ and the product of the Yukawa coupling $|y_d^{s4}y_d^{b4*}|$. This result is shown in the right figure of Fig.~\ref{Fig:ConstraintConv} where we use $v = 246$ GeV \cite{Patrignani:2016xqp}.
Since the constraint on $r_{sb}$ and $\theta_{sb}$ from the $\mathrm{Br}[B_s\rightarrow\mu^+\mu^-]_\mathrm{Exp}$ is stronger than that from the $\mathrm{Br}[\bar{B}\rightarrow X_s\gamma]_\mathrm{Exp}$ (See Fig.~\ref{Fig:ConstraintConv} left.),
we show the region with blue dots where $(M_4,|y_d^{s4}y_d^{b4*}|)$ satisfy the constraint from the $\mathrm{Br}[B_s\rightarrow\mu^+\mu^-]_\mathrm{Exp}$ and the quadrangle constraint Eq.~\eqref{Eq:UnitarityRelationbs2}.
One finds that the lower limit on the VLQ mass is around $5.5$ TeV for $\left|y_d^{s4}y_d^{b4*}\right|\sim 1$.

\begin{figure}[tbp]
  \begin{center}
 \begin{minipage}{0.46\hsize}
  \begin{center}
   \includegraphics[width=7.0cm]{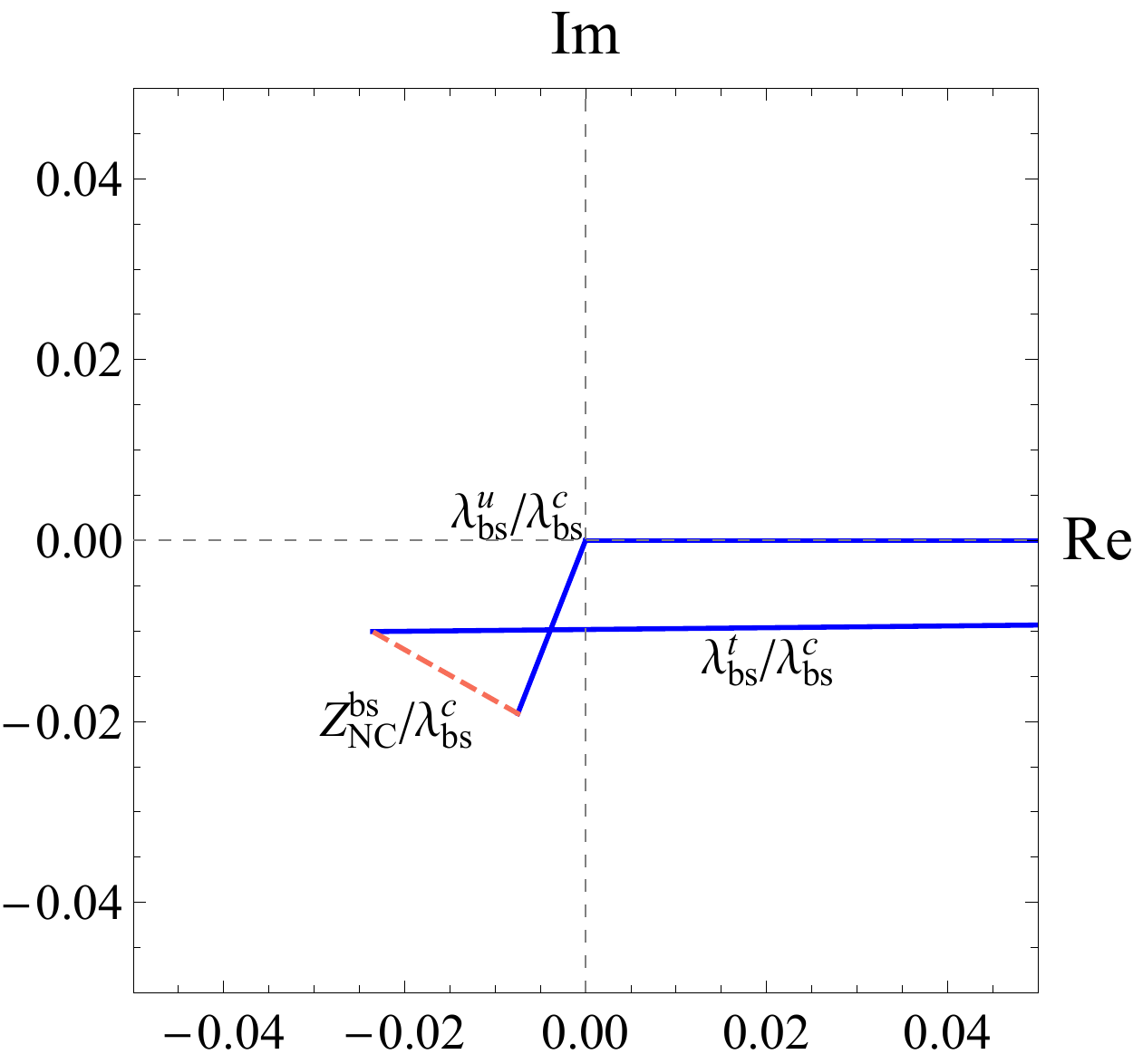}
  \end{center}
 \end{minipage}
 \hspace{1cm}
 \begin{minipage}{0.46\hsize}
  \begin{center}
   \includegraphics[width=7.0cm]{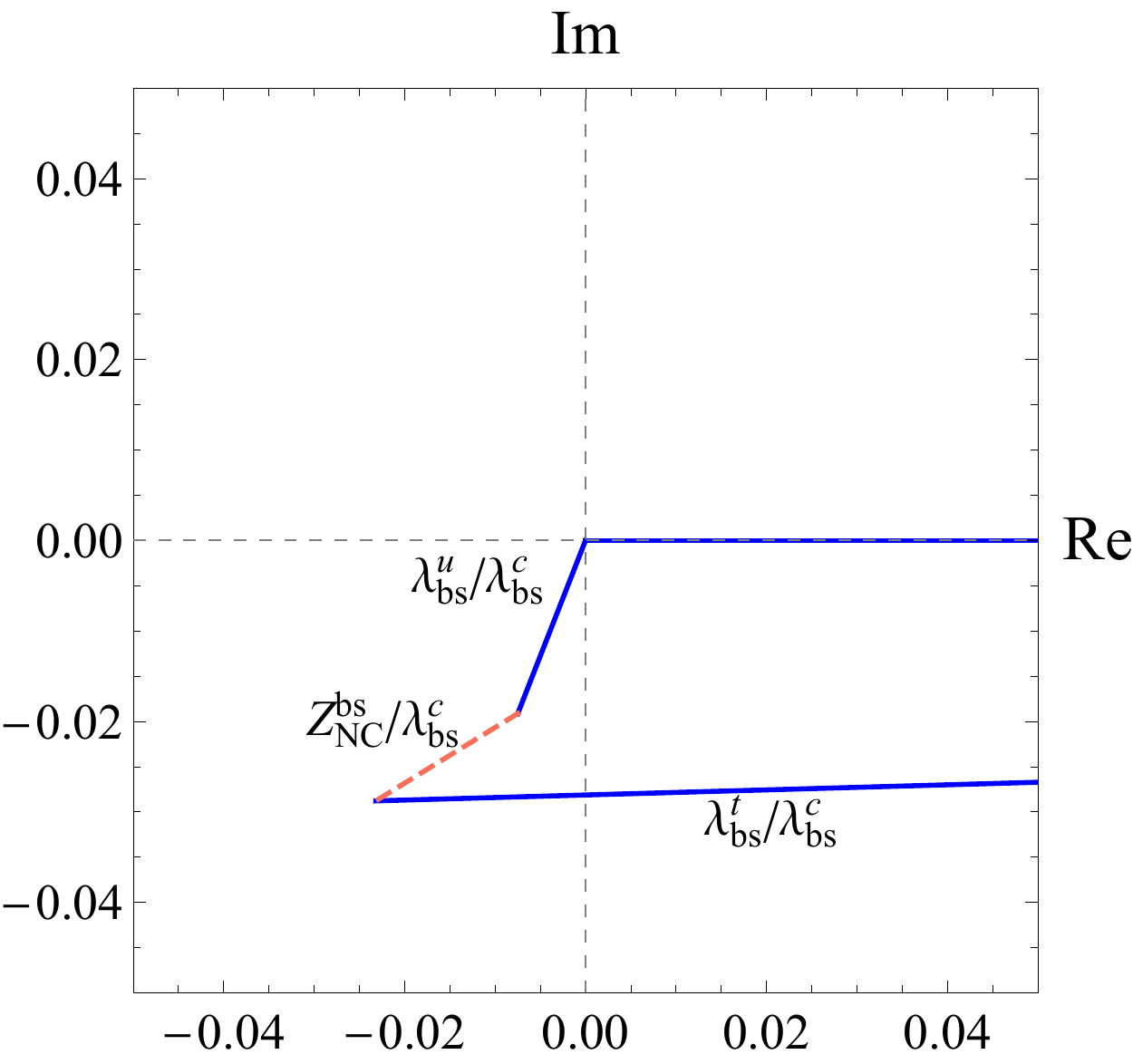}
  \end{center}
 \end{minipage}
 \caption{The violation of the CKM unitarity on the complex plane. In order to obtain these figures, we choose $(r_{sb},\theta_{sb})=(0.018,\pi/6)$ in the left figure and $(r_{sb},\theta_{sb})=(0.018,-\pi/6)$ in the right figure.}
 \label{Fig:Quadrangles}
 \end{center}
\end{figure}

Finally, we show the violation of the CKM unitarity on the complex plane in Fig.~\ref{Fig:Quadrangles}.
The definition of each side is the same as that of Fig.~\ref{Fig:Quadrangle}.
In order to obtain Fig.~\ref{Fig:Quadrangles}, we choose $r_{sb}=0.018$ and $\left|\theta_{sb}\right|=\pi/6$ and use the central values of the CKM matrix elements in Table~\ref{tab:ImputParameters}.
The side for $\lambda_{bs}^t$ is connected with the real axis at $(1,0)$.
The left figure is the case of $\theta_{sb}=\pi/6$ while the right figure is that of $\theta_{sb}=-\pi/6$.
One can see that the side for $Z_\mathrm{NC}^{sb}$ can be as large as that for $\lambda_{bs}^u$ and the sign of $\theta_{sb}$ affects the value of the angle $\beta_s$ in Fig.~\ref{Fig:Quadrangle}.

\section{Summary and Discussion}
\label{sec:5}

We studied the model which includes one down-type SU(2) singlet VLQ in addition to the SM quarks and showed the constraints on the model parameters from $\mathrm{Br}[B_s\rightarrow\mu^+\mu^-]$,  $\mathrm{Br}[\bar{B}\rightarrow X_s\gamma]$ and the quadrangle relation.
In order to analyze this model, we used the effective theory which is derived by integrating out the VLQ field.
We assume that the mass of the VLQ is much larger than the EW scale.
We matched the effective theory with the full theory not only at tree level but also at one-loop level and obtained the effective operators which are related to the radiative transition of the quarks. These operators correspond to the contribution from the diagrams including the VLQ in the internal line.
One can find that the coefficient of the photon dipole operator with the down-type quarks is consistent with the case of the full theory calculation given in Ref.~\cite{Handoko:1994xw}.

The other contributions to the radiative transitions come from the violation of the CKM unitarity and the diagrams which include the FCNC couplings among the SM quarks.
We obtained the effective Lagrangian for the $b\rightarrow s\gamma^{(*)}$ process arising from these contributions.

From our numerical results,
we obtained the constraints on the FCNC coupling $Z_\mathrm{NC}^{sb}$ and the new physics parameters $(r_{sb},\theta_{sb})$ defined in Eq.~\eqref{Eq.rtheta}.
We found that the dependence of $\mathrm{Br}[\bar{B}\rightarrow X_s\gamma]_\mathrm{VLQ}$ on $\theta_{sb}$ is weaker than that of  $\mathrm{Br}[B_s\rightarrow\mu^+\mu^-]_\mathrm{VLQ}$
and the constraint on the model parameters $r_{sb}$ and $\theta_{sb}$ from the $\mathrm{Br}[B_s\rightarrow\mu^+\mu^-]$ is more stringent than that from the $\mathrm{Br}[\bar{B}\rightarrow X_s\gamma]$.
One can discriminate the cases of different $|\theta_{sb}|$ through the $\mathrm{Br}[B_s\rightarrow\mu^+\mu^-]$ as shown in Fig.~\ref{Fig:ConstraintBsmumu}.
When the $|Z_\mathrm{NC}^{sb}|$ is order of $10^{-4}$,
the $\mathrm{Br}[B_s\rightarrow\mu^+\mu^-]$ becomes small (large) for $|\theta_{sb}|\simeq 0~(|\theta_{sb}|\simeq \pi)$ compared with that of the SM.
In Fig.~\ref{Fig:Quadrangles}, we showed the violation of the CKM unitarity on the complex plane when we chose $r_{sb}=0.018$ and $|\theta_{sb}|=\pi/6$.
The difference in the sign of the $\theta_{sb}$ affects the angle $\beta_s$, therefore we have to investigate the constraint from the observables related to the $\beta_s$ to further restrict the form of the violation of the CKM unitarity.

Although we focused on the case of $b\rightarrow s$ transitions,
the effective Lagrangian obtained in this paper can be applied to the FCNC transition for the other combinations of the down-type quarks.
The 4-Fermi operators in Eqs.~\eqref{Eq:OneLoopOpe_B}-\eqref{Eq:OneLoopOpe_G} and the effective Lagrangian for the off-shell photon contribute to $b\rightarrow s l^+l^-$ processes including $\bar{B}\rightarrow \bar{K^*}l^+l^-$.

Finally, we add a comment on the renormalization group (RG) effect.
One can not neglect the effect when the VLQ mass is much heavier than the EW scale.
When $M_4/M_W \sim 100$, one may expect about 10\% corrections to the Wilson coefficients.
Moreover, the expressions of the FCNC coupling, CKM matrix elements and down-type quark masses will be modified.
Including them, we will carry out the precise analysis elsewhere.


\vspace{0.5cm}
\noindent
{\bf Acknowledgement}

This work is supported by JSPS KAKENHI Grant Number JP16H03993 and JP17K05418 (T.M.).  This work is  also  supported  in  part  by  Grants-in-Aid  for  Scientific  Research  [No.  16J05332  (Y.S.), Nos.   24540272,  26247038,  15H01037,  16H00871,  and  16H02189  (H.U.)]  from  the  Ministry of Education, Culture, Sports, Science and Technology in Japan.


%

\end{document}